\documentclass[longauth]{aa}  

\usepackage[flushleft]{threeparttable}
\usepackage{graphicx}
\usepackage{amsmath}
\usepackage{url}
\usepackage[flushleft]{threeparttable}
\usepackage{booktabs}  
\usepackage{amssymb}
\usepackage[colorlinks=true,citecolor=blue, urlcolor=blue, linkcolor=blue]{hyperref}
\usepackage{multirow} 
\usepackage{times}
\usepackage{txfonts}
\usepackage{placeins}

\renewcommand{\v}{\ensuremath{\mathbf{v}}}

\makeatletter
\renewcommand*\aa@pageof{, page \thepage{} of \pageref*{LastPage}}
\makeatother

\begin{document}

\title{Exomoon search with VLTI/GRAVITY around the substellar companion HD 206893 B}\titlerunning{Exomoon search with GRAVITY}

\author{Q. Kral\inst{\ref{lesia}}\thanks{E-mail: quentin.kral@obspm.fr} \and J. Wang\inst{\ref{northwestern}} \and J. Kammerer\inst{\ref{esog}} \and S. Lacour\inst{\ref{lesia},\ref{esog}} \and M. Malin\inst{\ref{jhupa},\ref{stsci}} \and T. Winterhalder\inst{\ref{esog}} \and B. Charnay\inst{\ref{lesia}} \and C. Perrot\inst{\ref{lesia}} \and P. Huet\inst{\ref{lesia}} 
 \and R.~Abuter\inst{\ref{esog}}
 \and A.~Amorim\inst{\ref{lisboa},\ref{centra}}
 \and W.~O.~Balmer\inst{\ref{jhupa},\ref{stsci}}
 \and M.~Benisty\inst{\ref{ipag}}
 \and J.-P.~Berger\inst{\ref{ipag}}
 \and H.~Beust\inst{\ref{ipag}}
 \and S.~Blunt\inst{\ref{northwestern}}
 \and A.~Boccaletti\inst{\ref{lesia}}
 \and M.~Bonnefoy\inst{\ref{ipag}}
 \and H.~Bonnet\inst{\ref{esog}}
 \and M.~S.~Bordoni\inst{\ref{mpe}}
 \and G.~Bourdarot\inst{\ref{mpe}}
 \and W.~Brandner\inst{\ref{mpia}}
 \and F.~Cantalloube\inst{\ref{ipag}}
 \and P.~Caselli \inst{\ref{mpe}}
 \and G.~Chauvin\inst{\ref{cotedazur}}
 \and A.~Chavez\inst{\ref{northwestern}}
 \and A.~Chomez\inst{\ref{lesia},\ref{ipag}}
 \and E.~Choquet\inst{\ref{lam}}
 \and V.~Christiaens\inst{\ref{liege}}
 \and Y.~Cl\'enet\inst{\ref{lesia}}
 \and V.~Coud\'e~du~Foresto\inst{\ref{lesia}}
 \and A.~Cridland\inst{\ref{leiden}}
 \and R.~Davies\inst{\ref{mpe}}
 \and R.~Dembet\inst{\ref{lesia}}
 \and J.~Dexter\inst{\ref{boulder}}
 \and A.~Drescher\inst{\ref{mpe}}
 \and G.~Duvert\inst{\ref{ipag}}
 \and A.~Eckart\inst{\ref{cologne},\ref{bonn}}
 \and F.~Eisenhauer\inst{\ref{mpe}}
 \and N.~M.~F\"orster Schreiber\inst{\ref{mpe}}
 \and P.~Garcia\inst{\ref{centra},\ref{porto}}
 \and R.~Garcia~Lopez\inst{\ref{dublin},\ref{mpia}}
 \and T.~Gardner\inst{\ref{exeterAstro}}
 \and E.~Gendron\inst{\ref{lesia}}
 \and R.~Genzel\inst{\ref{mpe},\ref{ucb}}
 \and S.~Gillessen\inst{\ref{mpe}}
 \and J.~H.~Girard\inst{\ref{stsci}}
 \and S.~Grant\inst{\ref{mpe}}
 \and X.~Haubois\inst{\ref{esoc}}
 \and Th.~Henning\inst{\ref{mpia}}
 \and S.~Hinkley\inst{\ref{exeter}}
 \and S.~Hippler\inst{\ref{mpia}}
 \and M.~Houll\'e\inst{\ref{cotedazur}}
 \and Z.~Hubert\inst{\ref{ipag}}
 \and L.~Jocou\inst{\ref{ipag}}
 \and M.~Keppler\inst{\ref{mpia}}
 \and P.~Kervella\inst{\ref{lesia}}
 \and L.~Kreidberg\inst{\ref{mpia}}
 \and N.~T.~Kurtovic\inst{\ref{mpe}}
 \and A.-M.~Lagrange\inst{\ref{ipag},\ref{lesia}}
 \and V.~Lapeyr\`ere\inst{\ref{lesia}}
 \and J.-B.~Le~Bouquin\inst{\ref{ipag}}
 \and D.~Lutz\inst{\ref{mpe}}
 \and A.-L.~Maire\inst{\ref{ipag}}
 \and F.~Mang\inst{\ref{mpe}}
 \and G.-D.~Marleau\inst{\ref{duisburg},\ref{bern},\ref{mpia}}
 \and A.~M\'erand\inst{\ref{esog}}
 \and P.~Molli\`ere\inst{\ref{mpia}}
 \and J.~D.~Monnier\inst{\ref{umich}}
 \and C.~Mordasini\inst{\ref{bern}}
 \and D.~Mouillet\inst{\ref{ipag}}
 \and E.~Nasedkin\inst{\ref{mpia}}
 \and M.~Nowak\inst{\ref{lesia}}
 \and T.~Ott\inst{\ref{mpe}}
 \and G.~P.~P.~L.~Otten\inst{\ref{sinica}}
 \and C.~Paladini\inst{\ref{esoc}}
 \and T.~Paumard\inst{\ref{lesia}}
 \and K.~Perraut\inst{\ref{ipag}}
 \and G.~Perrin\inst{\ref{lesia}}
 \and O.~Pfuhl\inst{\ref{esog}}
 \and N.~Pourr\'e\inst{\ref{ipag}}
 \and L.~Pueyo\inst{\ref{stsci}}
 \and D.~C.~Ribeiro\inst{\ref{mpe}}
 \and E.~Rickman\inst{\ref{esa}}
 \and Z.~Rustamkulov\inst{\ref{jhueps}}
 \and J.~Shangguan\inst{\ref{mpe}}
 \and T.~Shimizu \inst{\ref{mpe}}
 \and D.~Sing\inst{\ref{jhupa},\ref{jhueps}}
 \and J.~Stadler\inst{\ref{mpa},\ref{origins}}
 \and T.~Stolker\inst{\ref{leiden}}
 \and O.~Straub\inst{\ref{origins}}
 \and C.~Straubmeier\inst{\ref{cologne}}
 \and E.~Sturm\inst{\ref{mpe}}
 \and L.~J.~Tacconi\inst{\ref{mpe}}
 \and A.~Vigan\inst{\ref{lam}}
 \and F.~Vincent\inst{\ref{lesia}}
 \and S.~D.~von~Fellenberg\inst{\ref{bonn}}
 \and F.~Widmann\inst{\ref{mpe}}
 \and J.~Woillez\inst{\ref{esog}}
 \and S.~Yazici\inst{\ref{mpe}}
 \and  the GRAVITY Collaboration
 \and K.~Abd El Dayem\inst{\ref{lesia}}
	\and	N.~Aimar\inst{\ref{porto},\ref{centra}}
	\and	A.~Berdeu\inst{\ref{esog},\ref{lesia}}
	\and	C.~Correia\inst{\ref{porto},\ref{centra}}
	\and	D.~Defr{\`e}re\inst{\ref{kul}}
	\and	M.~Fabricius\inst{\ref{mpe}}
	\and	H.~Feuchtgruber\inst{\ref{mpe}}    
	\and	A.~Foschi\inst{\ref{lesia}}          
	\and	S.F.~H{\"o}nig\inst{\ref{USouthampton}}    
	\and	S.~Joharle\inst{\ref{mpe}}
	\and	R.~Laugier\inst{\ref{kul}}
	\and	O.~Lai\inst{\ref{cotedazur}}
	\and	J.~Leftley\inst{\ref{USouthampton}}
	\and	B.~Lopez\inst{\ref{cotedazur}}
	\and	F.~Millour\inst{\ref{cotedazur}}
	\and	M.~Montarg{\`e}s\inst{\ref{lesia}}       
	\and	N.~Moruj{\~a}o\inst{\ref{porto},\ref{centra}}
	\and	H.~Nowacki\inst{\ref{cotedazur}}
	\and	J.~Osorno\inst{\ref{lesia}}
	\and	R.~Petrov\inst{\ref{cotedazur}}
	\and	P.O.~Petrucci\inst{\ref{ipag}}         
	\and	S.~Rabien\inst{\ref{mpe}}
	\and	S.~Robbe-Dubois\inst{\ref{cotedazur}}
	\and	M.~Sadun~Bordoni\inst{\ref{mpe}}
	\and	J.~S{\'a}nchez~Berm{\'u}dez\inst{\ref{UNAM}}
	\and	D.~Santos\inst{\ref{mpe}}
	\and	J.~Sauter\inst{\ref{mpia}}
	\and	J.~Scigliuto\inst{\ref{cotedazur}} 
	\and	F.~Soulez\inst{\ref{CRAL}}
	\and	M.~Subroweit\inst{\ref{cologne}} 
	\and	C.~Sykes\inst{\ref{USouthampton}}
	\and	the GRAVITY$^+$ Collaboration}

\institute{
LIRA, Observatoire de Paris, Université PSL, Sorbonne Université, Université Paris Cité, CY Cergy Paris Université, CNRS, 92190 Meudon, France
   \label{lesia} \and
   Center for Interdisciplinary Exploration and Research in Astrophysics (CIERA) and Department of Physics and Astronomy, Northwestern University, Evanston, IL 60208, USA
\label{northwestern}     \and
European Southern Observatory, Karl-Schwarzschild-Stra\ss e 2, 85748 Garching, Germany
\label{esog}     \and
   Department of Physics \& Astronomy, Johns Hopkins University, 3400 N. Charles Street, Baltimore, MD 21218, USA
\label{jhupa}      \and
   Space Telescope Science Institute, 3700 San Martin Drive, Baltimore, MD 21218, USA
\label{stsci}    \and
   Leiden Observatory, Leiden University, Einsteinweg 55, 2333 CC Leiden, The Netherlands
    \label{leiden}      \and
   Fakult\"at f\"ur Physik, Universit\"at Duisburg-Essen, Lotharstraße 1, 47057 Duisburg, Germany\label{duisburg}      \and
   Center for Space and Habitability, Universit\"at Bern, Gesellschaftsstrasse 6, 3012 Bern, Switzerland
\label{bern}      \and
   Max Planck Institute for Astronomy, K\"onigstuhl 17, 69117 Heidelberg, Germany
\label{mpia}         \and
   Universidade de Lisboa - Faculdade de Ci\^encias, Campo Grande, 1749-016 Lisboa, Portugal
\label{lisboa}      \and
   CENTRA - Centro de Astrof\' isica e Gravita\c c\~ao, IST, Universidade de Lisboa, 1049-001 Lisboa, Portugal
\label{centra}      \and
   Univ. Grenoble Alpes, CNRS, IPAG, 38000 Grenoble, France
\label{ipag}            \and
   Max Planck Institute for extraterrestrial Physics, Giessenbachstra\ss e~1, 85748 Garching, Germany
\label{mpe}      \and
   Université Côte d’Azur, Observatoire de la Côte d’Azur, CNRS, Laboratoire Lagrange, Bd de l'Observatoire, CS 34229, 06304 Nice cedex 4, France
\label{cotedazur}      \and
   Aix Marseille Univ, CNRS, CNES, LAM, Marseille, France
\label{lam}      \and
  STAR Institute, Universit\'e de Li\`ege, All\'ee du Six Ao\^ut 19c, 4000 Li\`ege, Belgium
\label{liege}      \and
   Department of Astrophysical \& Planetary Sciences, JILA, Duane Physics Bldg., 2000 Colorado Ave, University of Colorado, Boulder, CO 80309, USA
\label{boulder}      \and
   1st Institute of Physics, University of Cologne, Z\"ulpicher Stra\ss e 77, 50937 Cologne, Germany
\label{cologne}      \and
   Max Planck Institute for Radio Astronomy, Auf dem H\"ugel 69, 53121 Bonn, Germany
\label{bonn}      \and
   Universidade do Porto, Faculdade de Engenharia, Rua Dr.~Roberto Frias, 4200-465 Porto, Portugal
\label{porto}      \and
   School of Physics, University College Dublin, Belfield, Dublin 4, Ireland
\label{dublin}      \and
   Astrophysics Group, Department of Physics \& Astronomy, University of Exeter, Stocker Road, Exeter, EX4 4QL, United Kingdom
\label{exeterAstro}      \and
   Departments of Physics and Astronomy, Le Conte Hall, University of California, Berkeley, CA 94720, USA
\label{ucb}      \and
   European Southern Observatory, Casilla 19001, Santiago 19, Chile
\label{esoc}      \and
   Advanced Concepts Team, European Space Agency, TEC-SF, ESTEC, Keplerlaan 1, NL-2201, AZ Noordwijk, The Netherlands
\label{actesa}      \and%
   University of Exeter, Physics Building, Stocker Road, Exeter EX4 4QL, United Kingdom%
\label{exeter}\and%
   Astronomy Department, University of Michigan, Ann Arbor, MI 48109 USA
\label{umich}      \and
   Institute of Astronomy, University of Cambridge, Madingley Road, Cambridge CB3 0HA, United Kingdom
\label{cam}      \and
   Academia Sinica, Institute of Astronomy and Astrophysics, 11F Astronomy-Mathematics Building, NTU/AS campus, No. 1, Section 4, Roosevelt Rd., Taipei 10617, Taiwan
\label{sinica}      \and
   European Space Agency (ESA), ESA Office, Space Telescope Science Institute, 3700 San Martin Drive, Baltimore, MD 21218, USA
\label{esa}      \and
   Department of Earth \& Planetary Sciences, Johns Hopkins University, Baltimore, MD, USA
\label{jhueps}      \and
   Max Planck Institute for Astrophysics, Karl-Schwarzschild-Str. 1, 85741 Garching, Germany
\label{mpa}  \newpage    \and
   Excellence Cluster ORIGINS, Boltzmannstraße 2, D-85748 Garching bei München, Germany
\label{origins} \and
    Institute of Astronomy, KU Leuven, Celestijnenlaan 200D, 3001, Leuven, Belgium
\label{kul} \and 
    School of Physics \& Astronomy, University of Southampton, Southampton, SO17 1BJ, United Kingdom
\label{USouthampton} \and
    Instituto de Astronom{\'i}a, National Autonomous University of Mexico, Mexico City, Mexico
\label{UNAM} \and
    Univ. Lyon, Univ. Lyon 1, ENS de Lyon, CNRS, Centre de Recherche Astrophysique de Lyon UMR5574, F-69230, Saint Genis-Laval, France
\label{CRAL}       
}

   \date{Received September 15, 1932; accepted March 16, 1937}

  \abstract
   {Direct astrometric detection of exomoons remains unexplored. This study presents the first application of high-precision astrometry to search for exomoons around substellar companions.}
   {We investigate whether the orbital motion of the companion HD 206893 B exhibits astrometric residuals consistent with the gravitational influence of an exomoon or binary planet.}
   {Using the VLTI/GRAVITY instrument, we monitored the astrometric positions of HD 206893 B and c across both short (days to months) and long (yearly) timescales. This enabled us to isolate potential residual wobbles in the motion of component B attributable to an orbiting moon.}
   {Our analysis reveals tentative astrometric residuals in the HD 206893 B orbit. If interpreted as an exomoon signature, these residuals correspond to a candidate (HD 206893 B I) with an orbital period of approximately 0.76 years and a mass of $\sim$0.4 Jupiter masses. However, the origin of these residuals remains ambiguous and could be due to systematics. Complementing the astrometry, our analysis of GRAVITY $R=4000$ spectroscopy for HD 206893 B confirms a clear detection of water, but no CO is found using cross-correlation. We also find that AF Lep b, and $\beta$ Pic b are the best short-term candidates to look for moons with GRAVITY+.}
   {Our observations demonstrate the transformative
potential of high-precision astrometry in the search for exomoons, and proves the feasibility of the technique to detect moons with masses lower than Jupiter and potentially down to less than Neptune in optimistic cases. Crucially, further high-precision astrometric observations with VLTI/GRAVITY are essential to verify the reality and nature of this signal and attempt this technique on a variety of planetary systems.}


   \maketitle

\section{Introduction}\label{intro}

The detection of satellites around extrasolar planets or substellar companions, termed exomoons, remains an elusive frontier in modern astronomy. Moreover, there is no definition of what an exomoon is, and some ambiguity remains as to whether it may include, for instance, binary planets. Some have suggested that objects orbiting planets could be distinguished based on the center of mass of the system (planet+object). If the center of mass is outside the planet then it would be classified as a binary planet rather than an exomoon \citep{2002HiA....12..205S,2021ApJ...922L...2V}. Exomoons transiting isolated planetary-mass objects have also been discussed in \citet{2021ApJ...918L..25L}. Despite extensive observational efforts, no exomoon has been definitively confirmed to date. Numerous detection techniques have been actively pursued, though the intrinsic challenges of identifying such small, faint companions persist. The transit method represents the most prominent approach, leveraging distinctive light-curve signatures induced by planet-moon systems \citep{2007A&A...470..727S}. Complementary constraints may be derived from transit timing variations (TTVs) and transit duration variations (TDVs) \citep{1999A&AS..134..553S,2009MNRAS.392..181K,2009MNRAS.396.1797K}. Large-scale transit surveys have already imposed stringent limits on exomoon occurrence rates, constraining Galilean-analogue systems to fewer than 0.38 moons per planet for planets orbiting between $\sim$0.1 and 1.0 au \citep{2015ApJ...813...14K,2018AJ....155...36T}.

Alternative detection methodologies face distinct limitations. Microlensing surveys have not demonstrated significant promise for exomoon detection due to inherent smearing out of the exomoon signal \citep{2002ApJ...580..490H}. Radial velocity (RV) monitoring of planets themselves — rather than their host stars — has shown potential, yielding upper mass limits (e.g., $m \sin{i} = 2$ M$_{\rm Jup}$ for $<5$ day period moons around HR 8799 planets; \citealt{2021ApJ...922L...2V}). Recent advances combining high-contrast imaging with high-resolution spectroscopy now permit sensitivity to 1–4\% mass-ratio companions at separations comparable to the Galilean moons \citep{2023AJ....165..113R}. Nevertheless, detecting satellites with masses akin to those of Solar System moons (with a maximum satellite-to-planet mass ratio of around $10^{-4}$) remains beyond current capabilities.

To date, only a limited number of tentative exomoon candidates have emerged. The Jupiter-sized transiting planet Kepler-1625 b has drawn particular attention due to light-curve anomalies suggestive of a Neptune-mass satellite \citep{2018SciA....4.1784T}. However, the interpretation remains contested, with studies attributing the signal to an artifact of the data reduction, stellar activity, or unknown systematics \citep{2019A&A...624A..95H,2019ApJ...877L..15K}. Similarly debated is the proposed Neptune-sized companion to the super-Jupiter Kepler-1708 b \citep{2022NatAs...6..367K,2024NatAs...8..193H}. Other candidates include a wider-separation Jupiter-mass companion to the brown dwarf DH Tau B (projected separation: 10 au; mass ratio 0.1; \citealt{2020A&A...641A.131L}) and a 1.7 R$_\oplus$ object orbiting a free-floating planet \citep{2021ApJ...918L..25L}.

The scarcity of detections contrasts sharply with the ubiquity of moons in our Solar System ($>300$ confirmed), underscoring the observational challenges posed by sub-Earth-sized exomoons. Intriguingly, proposed candidates exhibit masses vastly exceeding those of solar system satellites – that have typical mass ratios $\sim 10^{-4}$ for Galilean moons –, suggesting initial discoveries may occupy the extreme end of the exomoon mass spectrum. This parallels early exoplanet surveys, which revealed unexpected populations such as hot Jupiters. Theoretical work now supports the plausibility of massive exomoons, whether via gravitational capture \citep{2019SciA....5.8665H} or formation within massive circumplanetary disks \citep{2020MNRAS.495.3763M}. Notably, moon masses may scale with host planet mass as M$_p^{3/2}$ \citep{2020ApJ...894..143B}, favoring searches around super-Jovian planets or brown dwarfs, as in this study.

Stability considerations further refine detection strategies. Exomoons must reside between their host’s Roche limit and $\sim$0.49 $R_{\rm Hill}$ (prograde orbits) or 0.93 $R_{\rm Hill}$ (retrograde orbits) \citep{2006MNRAS.373.1227D}. Close-in planets ($<1$ au) are poor candidates due to tidal destabilization \citep{2021PASP..133i4401D}, necessitating focus on wider-orbit systems. This motivates the application of high-precision astrometry, capable of probing distant planets.

In this work, we demonstrate the potential of the VLTI/GRAVITY instrument to detect exomoons via astrometric monitoring. GRAVITY’s tens of microarcsecond-level precision enables direct tracking of planetary orbital wobbles induced by massive satellites, probing periods spanning days to years – a regime where stable, detectable exomoons are most likely to reside. 

We present observations of the brown dwarf HD 206893 B \citep[mass $\sim$23 M$_{\rm Jup}$, semi-major axis $\sim$9 au, age $\sim$ 110 Myr;][but see our new fit that leads to a lower mass and larger semi-major axis]{2025AJ....169..175S}, a benchmark target for probing substellar companions. First identified with SPHERE \citep{2017A&A...597L...2M} around the F5V star HD 206893 \citep[distance: $\sim$ 40.8 pc,][]{2020yCat.1350....0G}, its properties remain debated: comparing the atmosphere properties from atmospheric models with evolutonary tracks suggest a possible younger age down to 3 Myr, and a planetary-mass regime as low as 5 M$_{\rm Jup}$ \citep{2021A&A...652A..57K}. Notably, HD 206893 B exhibits spectral characteristics of an L/T transition object, with extreme near-infrared colors indicative of high-altitude dust clouds \citep{2017A&A...608A..79D,2021AJ....161....5W,2021A&A...652A..57K}.

\begin{figure}
   \centering
   \includegraphics[width=\columnwidth]{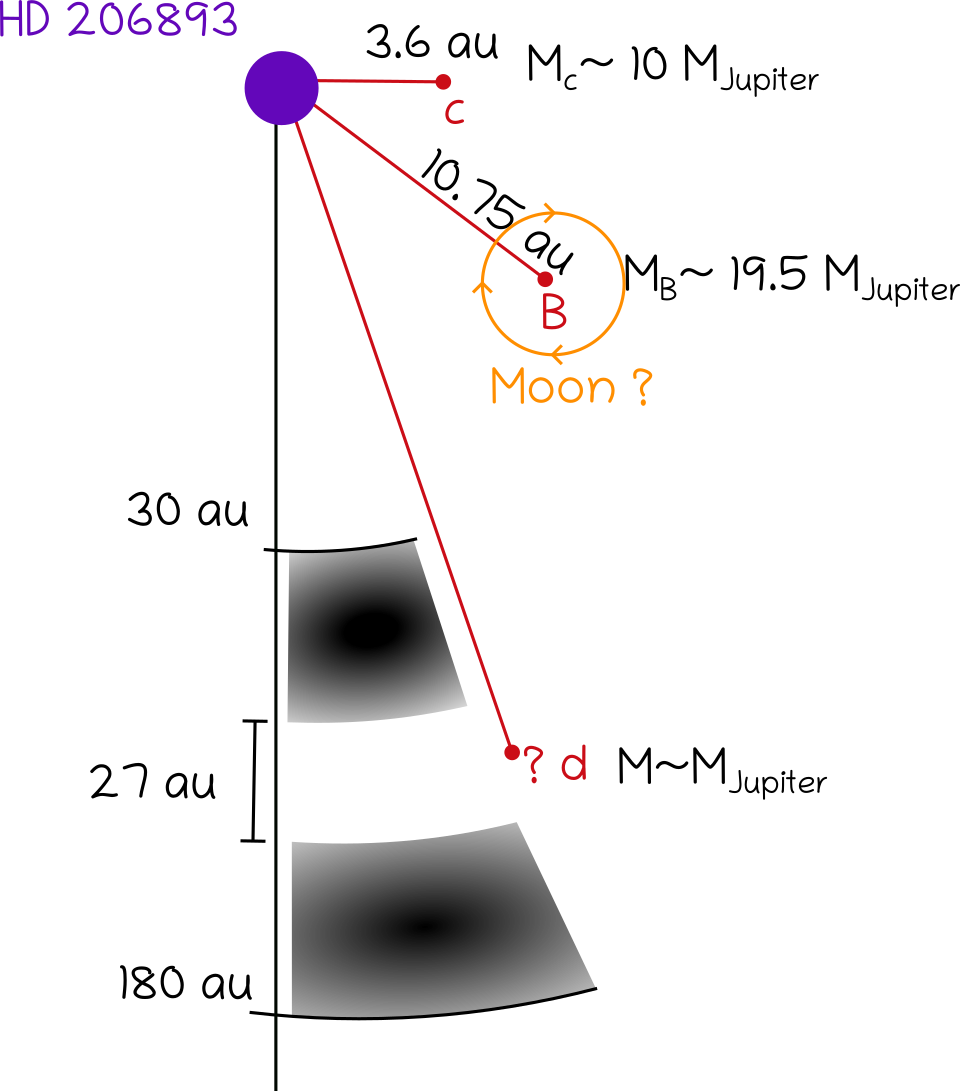}
   \caption{\label{figsystem} Schematic of the HD 206893 system.}
\end{figure}

The HD 206893 system hosts a complex architecture (see Fig.~\ref{figsystem}). A 
$\sim$10 M$_{\rm Jup}$ planetary-mass companion, HD 206893 c, orbits interior to the brown dwarf at $\sim$ 3.6 au \citep{Hinkley2023,2025AJ....169..175S}. ALMA observations further reveal a structured debris disk extending from 30 to 180 au, bisected by a 27 au gap \citep{2020MNRAS.498.1319M}. This gap may harbor an additional Jupiter-mass companion at $\sim$ 75 au, which, if confirmed, would constitute a third companion in the system.

Our analysis focuses on two key aspects: (1) quantifying GRAVITY’s sensitivity to exomoon-induced astrometric perturbations through precision monitoring of HD 206893 B’s orbital motion, and (2) conducting the first medium-resolution ($R=4000$) K-band spectral analysis of HD 206893 B’s atmosphere using GRAVITY data. By modeling the brown dwarf’s astrometric residuals, we establish detection limits for potential satellites and assess the instrument’s capability to resolve exomoon dynamical signatures. Moreover, we will refine the orbits of HD 206893 B and c using our new data. Simultaneously, the spectroscopic data may provide new insights into the atmospheric composition and cloud properties of this enigmatic L/T transition object.

 \section{Observations and data reduction}
 \label{sec:obs}

\begin{table}[ht]
\centering
\caption{GRAVITY astrometry for HD206893, B and c components.}
\begin{tabular}{lcccc}
\hline
\hline
& MJD & RA & DEC & $\rho$\tablefootmark{a}\\
\hline
B & 58681.392 & $130.75 \pm 0.06$ & $198.12 \pm 0.05$ & -0.57 \\
B & 58708.160 & $127.10 \pm 0.09$ & $199.20 \pm 0.11$ & -0.93 \\
B & 59453.093 & $20.06 \pm 0.07$ & $205.83 \pm 0.04$ & -0.68 \\
B & 60127.218 & $-79.30 \pm 0.06$ & $176.07 \pm 0.04$ & -0.37 \\
B & 60428.381 & $-120.91 \pm 0.08$ & $152.73 \pm 0.05$ & -0.07 \\
B & 60461.388 & $-125.29 \pm 0.05$ & $149.83 \pm 0.05$ & -0.32 \\
B & 60462.327 & $-125.47 \pm 0.08$ & $149.86 \pm 0.05$ & -0.73 \\
B & 60485.334 & $-128.32 \pm 0.14$ & $147.55 \pm 0.15$ & -0.66 \\
B & 60491.191 & $-129.33 \pm 0.23$ & $147.20 \pm 0.05$ & -0.65 \\
B & 60509.364 & $-131.50 \pm 0.04$ & $145.63 \pm 0.07$ & -0.57 \\
B & 60515.246 & $-132.26 \pm 0.03$ & $144.93 \pm 0.05$ & -0.40 \\
B & 60516.263 & $-132.40 \pm 0.03$ & $144.83 \pm 0.09$ & -0.13 \\
B & 60834.317 & $-170.50 \pm 0.03$ & $112.56 \pm 0.03$ & -0.48 \\
\hline
c & 59454.125 & $-52.76 \pm 0.12$ & $-69.36 \pm 0.13$ & -0.64 \\
c & 59485.110 & $-72.11 \pm 0.07$ & $-85.32 \pm 0.13$ & -0.85 \\
c & 59504.061 & $-69.30 \pm 0.06$ & $-86.73 \pm 0.07$ & -0.53 \\
c & 59721.403 & $-32.35 \pm 0.27$ & $-93.50 \pm 0.18$ & -0.87 \\
\hline
\hline
\end{tabular}
\tablefoot{\tablefoottext{a}{
$\rho$ stands for the Pearson correlation coefficient. The high correlation values are a direct consequence of the elongated geometry of the VLTI. }
}
\label{tab:astrometry}
\end{table}

HD~206893~B was observed with the GRAVITY instrument \citep{2017A&A...602A..94G} at the VLTI during eight epochs between April and July 2024 (Prog. ID 113.26D9.001). 
These observations complement those obtained as part of the Exo-GRAVITY large program \citep{2020SPIE11446E..0OL}. A final observation was conducted in June 2025 as part of the commissioning of the new adaptive optics system of GRAVITY+.
An observing log for all epochs is provided in Table~\ref{tab:obslog} in Appendix~\ref{app:log}. During the 2024 observing campaign, a short observational cadence was adopted to maximize sensitivity to moon-like companions around HD~206893~B. This strategy enhances the sensitivity to moons orbiting the brown dwarf companion on timescales of days to months.

All data were obtained with the four 8.2 m Unit Telescopes (UTs) in dual-field on-axis mode. In this configuration, a 50/50 beamsplitter directed half of the starlight to the GRAVITY fringe tracker, while the other half (either from the star or the companion) was sent to the GRAVITY science spectrometer, which was operated at a spectral resolving power of $R \sim 4000$ in the K-band ($\sim2$--$2.4~\mu$m). This setup allowed for the calibration of the companion exposures using the host star exposures, eliminating the need for dedicated calibrator observations.

The data were reduced using the ESO GRAVITY pipeline version 1.6.4b and further processed with the GRAVITY Consortium Python tools\footnote{Available for download at \url{https://gitlab.obspm.fr/mnowak/exogravity}} to extract relative astrometry and \textit{K}-band contrast spectra (relative to the host star) of HD~206893~B \citep{2014SPIE.9146E..2DL}. The data reduction followed the procedure described in Appendix~A of \citet{2020A&A...633A.110G}. A recent improvement was achieved by down-weighting baselines that have a high degree of self-subtraction caused by post-processing. This allowed for a more aggressive suppression of the stellar flux by increasing the order of the polynomial fit, without excessively amplifying systematic noise. Reductions were therefore performed using different polynomial orders for the post-processing \citep[see the impact of polynomial order in][]{2025A&A...694A.277L}, specifically orders 3, 4, and 6. The resulting astrometry is listed in Table~\ref{tab:pos}. All the datasets were uniformly reduced to give the astrometry adopted in this paper, summarized in Table~\ref{tab:astrometry}.

\begin{figure}
   \centering
   \includegraphics[width=\columnwidth]{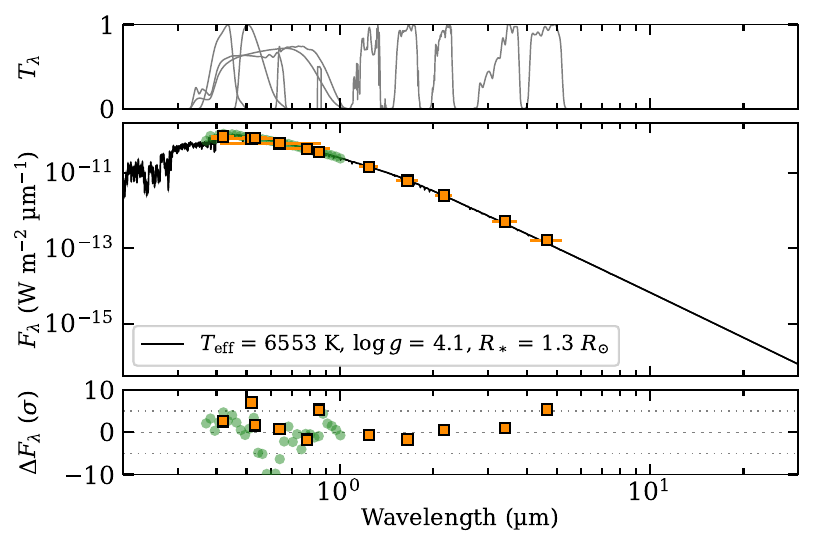}
   \caption{\label{figspectrumhost} Model spectrum of HD~206893~A obtained by fitting a BT-NextGen stellar model atmosphere (black line) to archival photometry (orange points) and the Gaia XP spectrum (green points). The top panel shows the filter transmission curves and the bottom panel shows the residuals between data and model.}
\end{figure}

To convert the contrast spectra into flux-calibrated spectra, we multiplied them by a model spectrum of the host star HD~206893~A. For this purpose, we fitted a BT-NextGen stellar atmosphere model \citep{2012RSPTA.370.2765A} to archival photometry from Gaia, 2MASS, Tycho, and WISE, as well as the Gaia XP-spectrum \citep{2021A&A...652A..86C,2023A&A...674A...2D} of HD~206893~A. Interstellar extinction was not included in the fit, as the Gaia DR3 catalog reports a \textit{G}-band extinction of 0~mag \citep{2023A&A...674A...1G}. The surface gravity was fixed to the Gaia DR3 value of $\log g = 4.13$. The fit yields an effective temperature of $6554^{+29}_{-27}$~K, consistent within 1$\sigma$ with the value reported by \citet{2017A&A...608A..79D}. Figure~\ref{figspectrumhost} displays the resulting model spectrum for HD~206893~A along with the archival photometry used in the fit, providing a good fit in band K.

\begin{figure*}
   \centering
   \includegraphics[width=18.cm]{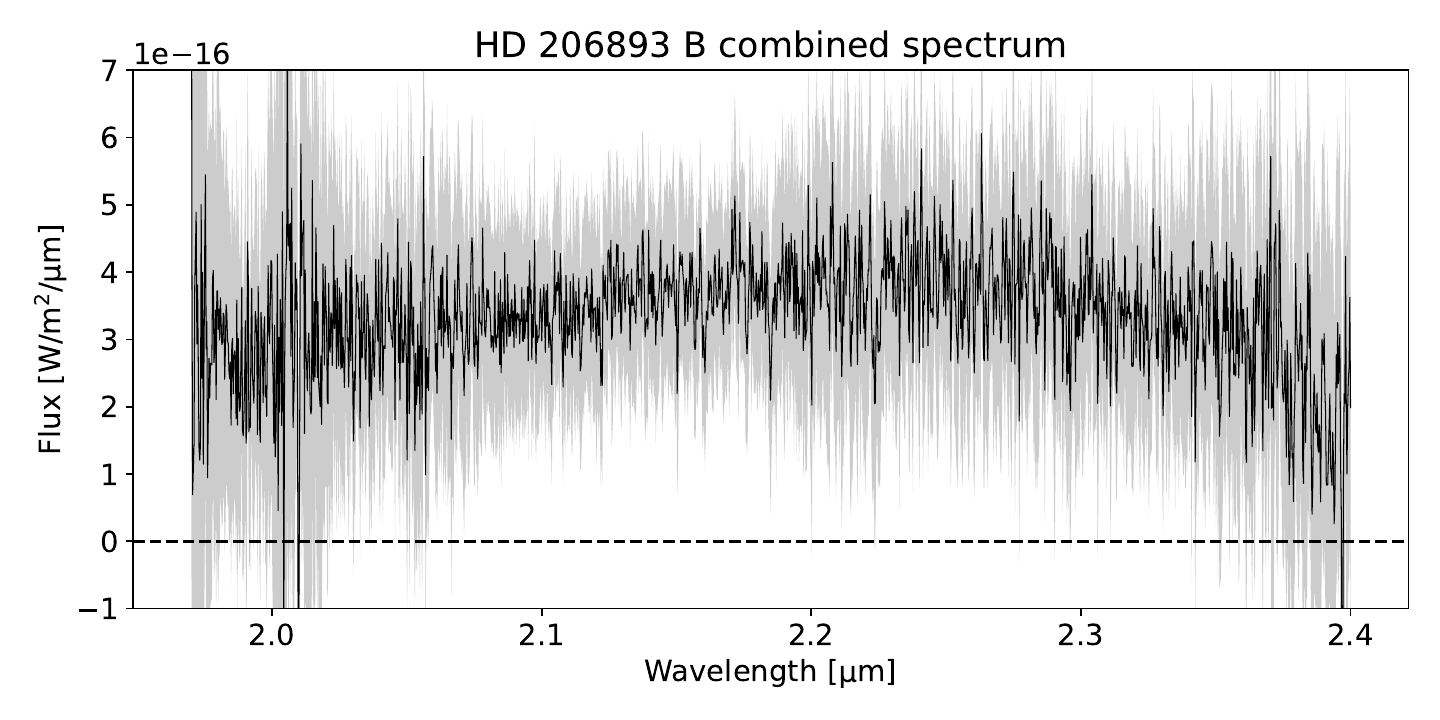}
   \caption{\label{figspectrum} Combined R $\sim$ 4000 spectrum obtained with GRAVITY for HD 206893 B by averaging the data obtained at different epochs (see Fig.~\ref{figspectrumapp}).}
\end{figure*}

The companion flux spectra are then obtained by multiplying the GRAVITY contrast spectrum with the stellar model spectrum. We propagate the uncertainties of the stellar model and the correlations of the GRAVITY spectra throughout the entire process. This yields a companion flux spectrum for each epoch. To obtain the highest possible signal-to-noise ratio (SNR), the individual epochs are then averaged together to a final companion flux spectrum by computing a covariance-weighted mean spectrum. Before, though, we applied scaling factors to each individual epoch to bring them into better agreement with each other. The best fitting scaling factors are shown in Figure~\ref{figspectrumapp}. The scaling factors were introduced to mitigate the $\sim$10\% systematic flux variability between epochs caused by varying PSF shape due to atmospheric turbulence and instrumental effects. The final companion flux average spectrum is shown in Figure~\ref{figspectrum}.

 \section{Exomoon search and new predictions for the orbits of HD 206893 B and c}
 \subsection{Back of the envelope predictions for the moon's astrometric effect}\label{back}

 Let us consider a moon of mass $M_{\rm moon}$ orbiting a planet of mass $M_{\rm pla}$. They will both orbit around their barycenter, with the semi-major axis of the planet’s orbit due to the exomoon given by
 
\begin{equation}
a_{\textrm{ast}}=a_{\textrm{moon}} \frac{M_{\textrm{moon}}}{M_{\textrm{pla}}+M_{\textrm{moon}}} \approx q \, a_{\textrm{moon}}, 
\end{equation}

\noindent where $a_{\textrm{moon}}$ is the distance between the planet and the moon, and $q$ is the moon-to-planet mass ratio $M_{\textrm{moon}}/M_{\textrm{pla}}$.

\begin{figure}
   \centering
   \includegraphics[width=\columnwidth]{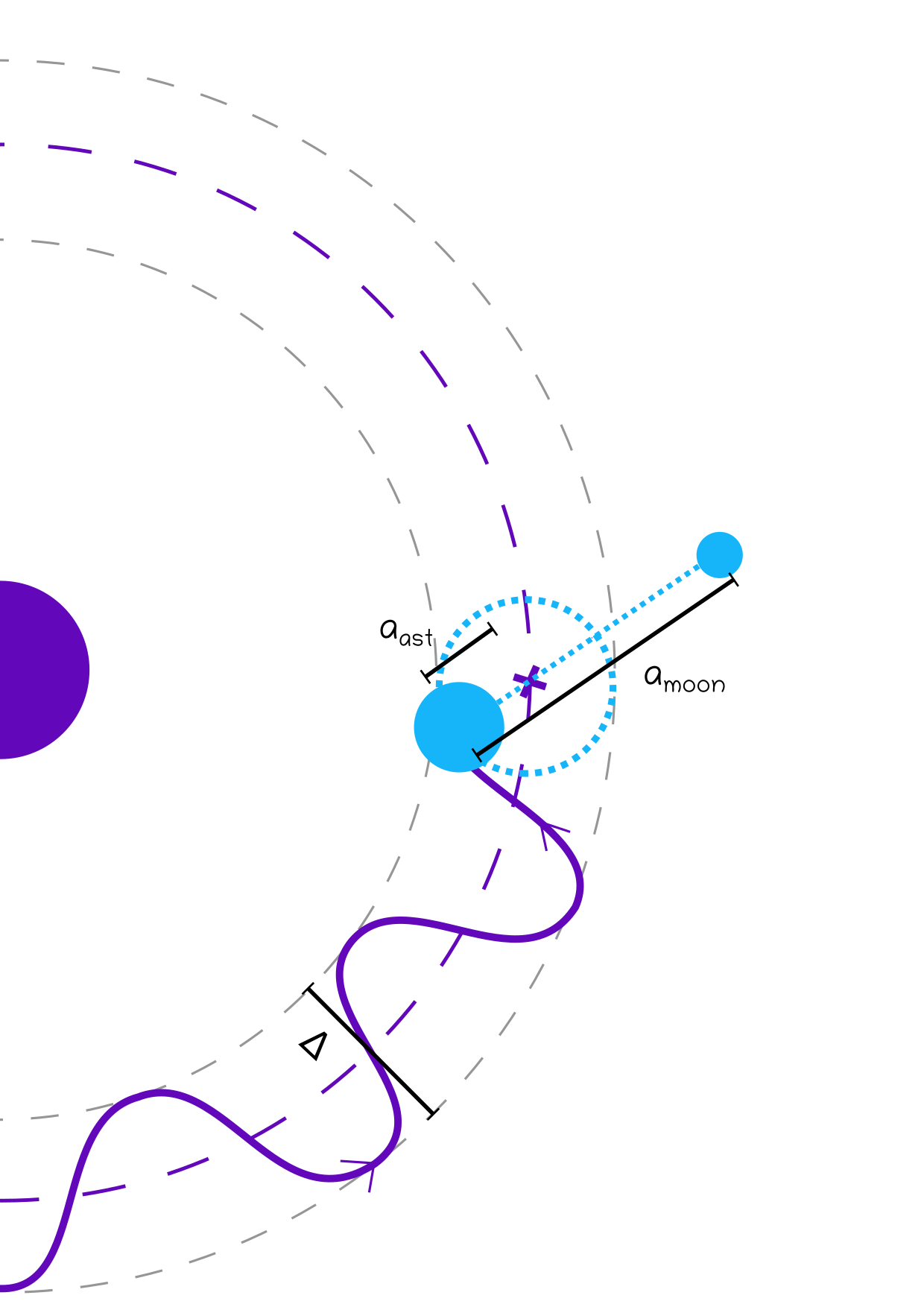}
   \caption{\label{figbary} Diagram of the orbit of HD 206893 B with a potential moon around it, causing it to shift slightly due to the moon's gravitational potential. It illustrates the notation used in section \ref{back}.}
\end{figure}

If we define the maximum astrometric angular displacement $\Delta$ for a full orbit of the planet around its barycenter due to the moon's gravity, we get 

\begin{equation}
 \Delta=\frac{2 a_{\textrm{ast}}}{d},
\end{equation}

\noindent where $d$ is the system's distance to the Sun (see Fig.~\ref{figbary}). Then, we can readily derive

\begin{equation}
 q=\left[\frac{2 a_{\textrm{moon}}}{d \Delta}-1\right]^{-1}.
\end{equation}

Note that this derivation assumes that we measure the wobble of the giant planet or companion, while we actually measure the wobble of the center of light of the pair with GRAVITY \citep{2016A&A...592A..39K}. The moon may have a non-negligible flux contribution depending on its albedo in the infrared compared to the host. However, it is neglected in this study, given uncertainties, and that we consider moons with much lower masses than the moon host. 

If the orbit has an eccentricity $e$, then the minimum astrometric displacement seen on the sky would be twice the semi-minor axis $b_{\textrm{ast}}=a_{\textrm{ast}} \sqrt{1-e^2}$ so that $\Delta=2 a_{\textrm{ast}} \sqrt{1-e^2}/d$. In this case, we obtain \citep{2023AJ....165..113R}

\begin{equation}
 q=\left[\frac{2 a_{\textrm{moon}} \sqrt{1-e^2}}{d \Delta}-1\right]^{-1}.
\end{equation}

\begin{figure}
   \centering
   \includegraphics[width=9.cm]{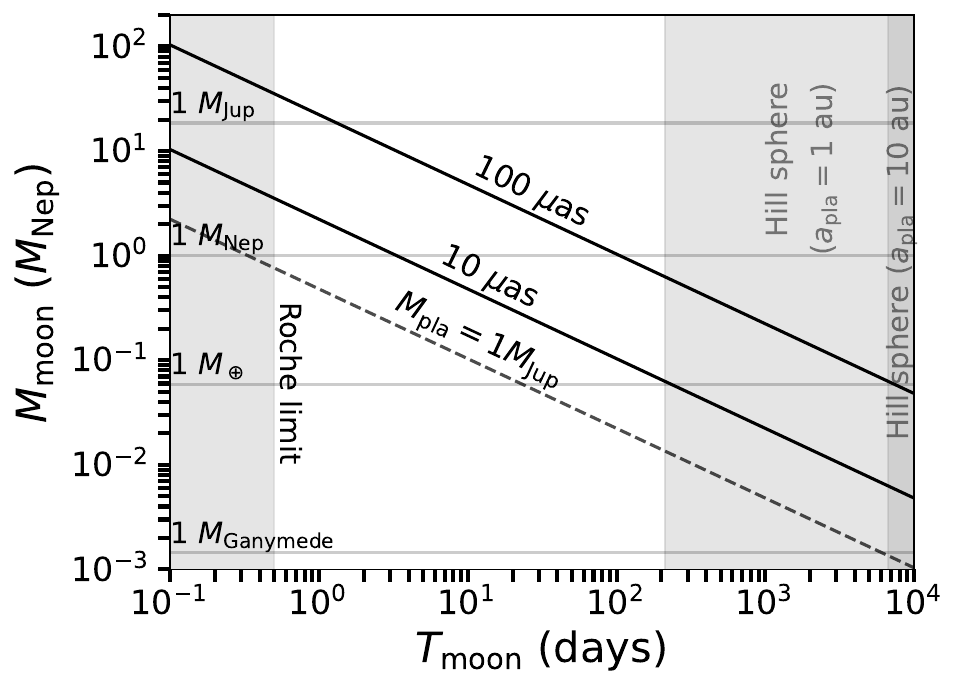}
   \caption{\label{figanalgen} Analytical predictions of $M_{\rm moon}$ (the mass of the moon in masses of Neptune) as a function of $T_{\rm moon}$ (the period of the moon in days). The solid black lines show the model's prediction for astrometric accuracies of 10 and 100 ${\rm \mu}$as, as indicated on the graph. The values used are those of Eq.~\ref{Mmoon}, i.e. $d=10$ pc and $M_{\rm pla}=10 \, {\rm M}_{\rm Jup}$. The dotted line shows the case of 10 $\mu$as considering $M_{\rm pla}=1 \, {\rm M}_{\rm Jup}$. If the system is instead at 100 pc, the lines are pushed up by a factor of 10. The areas beyond the Roche limit and the Hill sphere are shown in grey. The Hill sphere is shown for two values of the semi-major axis of the planet $a_{\rm pla}$ equal to 1 (closest) and 10 au (farthest). The horizontal lines indicate moon masses of 1 M$_{\rm Jup}$, 1 M$_{\rm Nep}$, 1 M$_\oplus$, and 1 M$_{\rm Ganymede}$ from top to bottom.}
\end{figure}

For simplicity, assuming that the eccentricity is zero and that $q \ll 1$, then we obtain

\begin{equation}\label{eqast}
 q=\frac{1}{2} \left(\frac{a_{\textrm{moon}}}{\textrm{au}}\right)^{-1}\left(\frac{d}{\textrm{pc}}\right)\left(\frac{\Delta}{\textrm{1''}}\right),
\end{equation}

\noindent or converting to the period $T_{\rm moon}$ of the orbit using Kepler's third law, we finally obtain

\begin{equation}\label{Mmoon}
 \frac{M_{\rm moon}}{{\rm M}_{\rm Nep}}=0.48\, \left(\frac{T_{\textrm{moon}}}{\textrm{10 d}}\right)^{-2/3}\left(\frac{d}{\textrm{10 pc}}\right)\left(\frac{\Delta}{10 \, {\rm \mu} \textrm{as}}\right)\left(\frac{M_{\textrm{pla}}}{10 \, {\rm M}_{\rm Jup}}\right)^{2/3},
\end{equation}

\noindent where M$_{\rm Nep}$ is Neptune's mass equal to 17.15 M$_\oplus$, and M$_{\rm Jup}$ the Jupiter's mass equal to 317.8 M$_\oplus$. In the formula above, we use units of day, parsec, and microarcsec for the period, distance, and astrometric signature, respectively. This equation shows that exomoons with masses around that of Neptune may be targeted with an astrometry at 10 ${\rm \mu}$as level for the closest systems. It may even go down to an Earth mass for moons located at larger periods and down to the mass of the Jupiter moon Ganymede of mass M$_{\rm Ganymede} \sim 0.025$ M$_\oplus$ around less massive planets than assumed here (10 M$_{\textrm{Jup}}$) located at larger distances from the central star with their Hill spheres being much larger, allowing for stable moons at several year periods. To better visualize this equation, we plot it in Fig.~\ref{figanalgen}.

As stated earlier, the moon should remain within the planet's Hill sphere $R_{\rm Hill}=a_{\rm pla}(1-e_{\rm pla})(M_{\rm pla}/(3 M_\star))^{1/3}$, where $a_{\rm pla}$ is the planet's semi-major axis, $e_{\rm pla}$ the planet's eccentricity, and $M_\star$ the stellar mass. The moon should also be located beyond the Roche limit \citep{roche1849} defined as $R_{\rm roche}=R_{\rm pla} \, (4 \pi \rho_{\rm pla}/(\gamma \rho_{\rm moon}))^{1/3}$, where $R_{\rm pla}$ is the planet's radius, $\rho_{\rm pla}$ the planet's density, $\rho_{\rm moon}$ the moon's density, and $\gamma$ is a dimensionless geometrical parameter with $\gamma=4\pi/3$ for a sphere, but is smaller for an object that takes a non-spherical shape \citep{1999ssd..book.....M}. \citet{2007Sci...318.1602P} find that $\gamma=1.6$ may be more realistic for moonlets. In the case of an incompressible fluid, taking full account of the feedback between the moon's deformed shape and its gravity field smoothes the cusps and elongates the moon further \citep{1969efe..book.....C,1999ssd..book.....M}, leading to an even smaller value, $\gamma \approx 0.85$. This sets some limits on the moon's period such that (within the Hill sphere)

\begin{equation}
 \frac{T_{\rm moon}}{\textrm{yr}} < 18\, \left(\frac{M_\star}{{\rm M}_\odot}\right)^{-1/2}\left(\frac{a_{\rm pla}}{\textrm{10 au}}\right)^{3/2}\left(1-e_{\rm pla}\right)^{3/2},
\end{equation}

\noindent and (beyond the Roche limit)

\begin{equation}
 \frac{T_{\rm moon}}{\rm d} > 0.5\, \left(\frac{\gamma}{1}\right)^{-1/2}\left(\frac{\rho_{\rm moon}}{1000 \,\textrm{kg/m}^3}\right)^{-1/2},
\end{equation}

\noindent leading to typical moon periods between a day and several decades, but note the strong dependence on the planet semi-major axis $a_{\rm pla}$, which can push the Hill radius to 210 day period for $a_{\rm pla}=1$ au instead of 18 yr for 10 au, or 580 yr for 100 au. 

\begin{figure}
   \centering
   \includegraphics[width=9.cm]{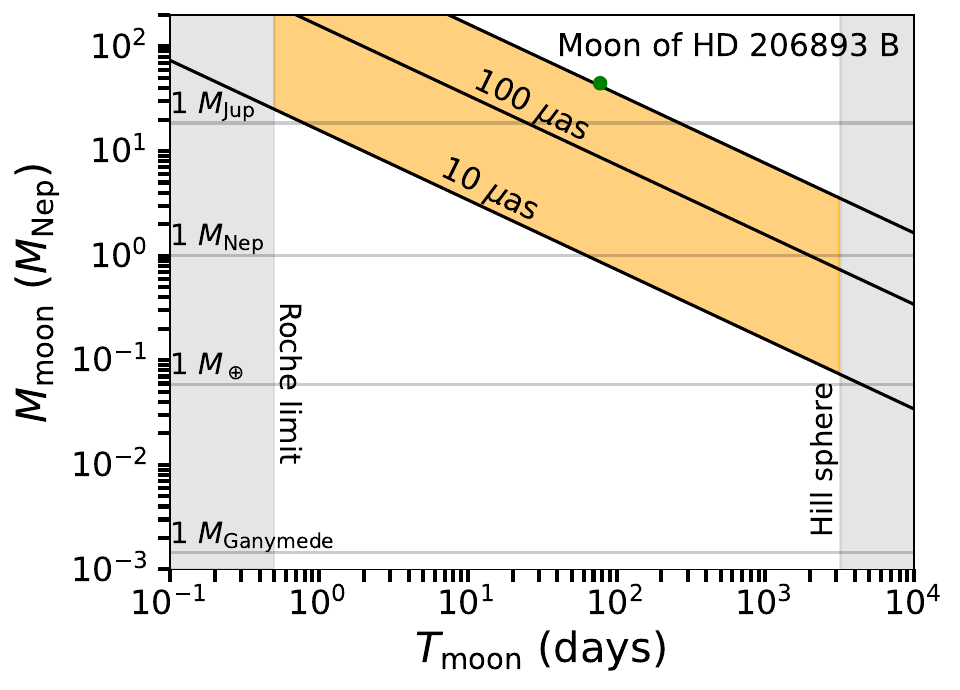}
   \caption{\label{figanal} Analytical predictions for the astrometric effect of a moon with varying mass and distance to HD 206893 B. GRAVITY could in principle detect moons lower than a Neptune mass at larger periods. The fit of the GRAVITY data is shown as a green point assuming the mass and period as upper limits in our tentative detection. The orange area shows the plausible parameter space where a moon could be located, and still be potentially detectable with GRAVITY assuming an astrometric accuracy of 10 $\mu$as (see section \ref{fitmoon}).}
\end{figure}

Now we focus on the HD 206893 system and produce a prediction for the presence of an exomoon around HD 206893 B using its specific set of parameters. Using that $d=40.8$ pc, $M_{\rm pla}=19.5 $ M$_{\rm Jup}$ (see next section), we plot the moon mass model prediction as a function of moon period in Fig.~\ref{figanal}. We also adapt the Hill radius by using $M_\star=1.31$ M$_\odot$, $a_{\rm pla}=10.75$ au, and $e_{\rm pla}=0.07$ (see our fit in the next section). In addition, we show a green point obtained by fitting the GRAVITY data that represents upper limits in moon mass and period derived in section \ref{fitmoon} (i.e. an upper limit of 0.8 M$_{\rm Jup}$ in mass and using the tentative detection semi-major axis of 0.22 au or 278 days). The orange area represents the allowed parameter space of a plausible moon that would hide around HD 206893 B but would still be detectable with GRAVITY given an astrometric accuracy of 10 $\mu$as. We note that the orange zone extends slightly beyond the analytical predictions because our upper limit accounts for different orientations of the system (e.g., obliquity, inclination), whereas the analytical model does not.

\subsection{Refining the orbit of HD 206893 B and c}
We repeat the orbit modeling following the procedure described in \citet{Hinkley2023} to obtain updated orbital posteriors for HD 206893 B and c. We fit the relative astrometry, Hipparcos-Gaia absolute astrometry from the EDR3 edition of the HGCA \citep{Brandt2021}, and stellar radial velocity data \citep{Grandjean2019, Hinkley2023} using \texttt{orbitize!} \citep{Blunt2020, Blunt2024}. We did not include the companion radial velocity of B measured in \citet{2025AJ....169..175S}, as it did not provide any additional constraints to the orbit fit. 

We used the parallel-tempered affine-invariant sampler \texttt{ptemcee} \citep{Foreman-Mackey2013, Vousden2016} with 20 temperatures and 1000 walkers per temperature to sample the posterior. We discarded the first 35,000 steps as ``burn-in" after visual inspection of the walkers, using only the last 5000 steps from the lowest temperature walkers to construct the posterior. We present the updated orbital parameters in Table \ref{table:detailed_orbit}. 

In broad strokes, the orbital configuration of the two planets remains the same as from \citet{Hinkley2023}. The substantial increase in GRAVITY astrometry of B has resulted in a much more precise orbit solution for both planets, since the visual orbit of B also contains information on the orbit of c (similar to GRAVITY observations of $\beta$ Pic b and c; \citealt{Lacour2021}). The orbit of HD 206893 B is nearly circular, and with a 1 au larger semimajor axis than what was found in \citet{Hinkley2023}. This results in a dynamical mass for B that is 8 M$_\textrm{Jup}$ lower than what was found in \citet{Hinkley2023}. The orbit of HD 206893 c is also more circular ($0.283^{+0.019}_{-0.017}$ compared to $0.41^{+0.03}_{-0.03}$) and is now nearly coplanar with HD 206893 B. This is a much more dynamically stable configuration than the nearly unstable orbital solution found in \citet{Hinkley2023}. The epicyclic motion in the visual orbit of B due to the orbit of c is now detected in the GRAVITY data, which helps constrain the orbit of c, which previously relied heavily on the stellar RV and astrometry.

\begin{table}[ht]
    \centering
    \caption{Orbital fit of the B and c components in the HD 206893 system.}
    \begin{tabular}{ccc}
    \hline
 Quantity  & Prior & Posterior \\
    \hline
    \hline
 $a_B$ (au)               & LogUniform(1, 100) & $10.75 \pm {0.08}$  \\
 $e_B$                    & Uniform(0, 1) & $0.069^{+0.011}_{-0.012}$  \\
 $i_B$ (\degr)            & $\sin(i)$ & $142.0^{+0.4}_{-0.5}$  \\
 $\omega_B$ (\degr)       & Uniform(0, 2$\pi$) & $102^{+7}_{-5}$ \\
 $\Omega_B$ (\degr)       & Uniform(0, 2$\pi$) & $63.8^{+0.5}_{-0.6}$ \\
 $\tau_B$                 & Uniform(0, 1) & $0.133^{+0.020}_{-0.013}$ \\
 \hline
 $a_c$ (au)               & LogUniform(1, 100) & $3.74^{+0.02}_{-0.03}$ \\
 $e_c$                    & Uniform(0, 1) & $0.283^{+0.019}_{-0.017}$ \\
 $i_c$ (\degr)            & $\sin(i)$ & $142.5^{+0.9}_{-0.8}$ \\
 $\omega_c$ (\degr)       & Uniform(0, 2$\pi$) & $27 \pm 5$ \\
 $\Omega_c$ (\degr)       & Uniform(0, 2$\pi$) & $65.4^{+3.6}_{-3.5}$ \\
 $\tau_c$                 & Uniform(0, 1) & $0.694^{+0.009}_{-0.010}$ \\
 \hline
 Parallax (mas)         & $\mathcal{N}$(24.5275, 0.0354) & $24.5252^{+0.0350}_{-0.0347}$ \\
 $\gamma_{RV}$ (m/s)         & Uniform($-5000$, 5000) & $146^{+8}_{-7}$ \\
 $\sigma_{RV}$ (m/s)         & LogUniform(0.1, 100) & $43^{+4}_{-3}$ \\
 $M_{B}$ (M$_\textrm{Jup}$)  & Uniform(1, 50) & $19.5^{+1.4}_{-1.3}$ \\
 $M_{c}$ (M$_\textrm{Jup}$)  & Uniform(1, 50) & $11.1^{+0.4}_{-0.5}$ \\
 $M_{*}$ (M$_\odot$)  & $\mathcal{N}$(1.29, 0.10) & $1.31^{+0.03}_{-0.02}$ \\
 \hline
 \multicolumn{3}{c}{Derived Parameters} \\
 \hline
 $P_B$ (yr)         & -- & $30.5^{+0.6}_{-0.7}$ \\ 
 $P_c$ (yr)         & -- & $6.28^{+0.08}_{-0.07}$ \\ 
 Mutual Inc.\ (\degr)    & -- & $1.8^{+1.5}_{-0.7}$ \\ 
 \hline
 \end{tabular}
\tablefoot{Orbital parameters in Jacobi coordinates. $a$, $e$, $i$, $\omega$, $\Omega$, and $\tau$ are the semi-major axis, eccentricity, inclination, argument of periapsis, longitude of ascending node, and the relative epoch of periastron, respectively. They are shown for the B and c components with associated subscripts. Then we list the parallax, the stellar RV offset ($\gamma_{RV}$), the stellar RV jitter ($\sigma_{RV}$), the mass of the B and c components, and that of the star. We finish with the orbital periods of the B and c components and their mutual inclinations.}
 \label{table:detailed_orbit}
\end{table}

\subsection{Fit of astrometric residuals to look for an exomoon}\label{fitmoon}

To search for a possible companion orbiting HD 206893 B, we performed a 3-body hierarchical fit with \texttt {orbitize!}. In this hierarchical model, we assume a companion is orbiting B, and that the barycenter of these two bodies orbits around the star. This is implemented in practice in \texttt{orbitize!} by setting B to be body 0, the potential companion to B to be body 1, and the host star HD 206893 to be body 2, and by setting a non-zero mass for B (19.5 M$_\textrm{Jup}$; Table \ref{table:detailed_orbit}). In addition to the gravitational influence of the host star and any potential moon, the orbit of B is also influenced by c. This is not captured by the hierarchical model implemented in \texttt{orbitize!}. Thus, we subtract out the perturbations in the visual orbit of B due to the gravitational influence of c (i.e., c causing the star to move about the system barycenter, which is encoded in the relative astrometry between B and the host star; see \citealt{Lacour2021}) using the maximum posterior probability orbit from the previous section. 

For simplicity, we assumed the potential orbital companion to HD 206893 B is on a circular orbit. We also set priors on the mass of the potential companion to be between $10^{-3}$ and 10~$M_\textrm{Jup}$ and the semimajor axis to be between 0.001 and 1 AU. We did not place any constraints on the orbital orientation of the moon relative to the orbit of B. 
We used \texttt{ptemcee} to sample the orbital posterior using 20 temperatures and 1000 walkers per temperature. We discarded the first 10,000 steps of each walker as burn-in after visual inspection, and used the remaining 10,000 steps of each lowest-temperature walker to construct the posterior.

\begin{figure}[ht]
   \centering
   \includegraphics[width=9.cm]{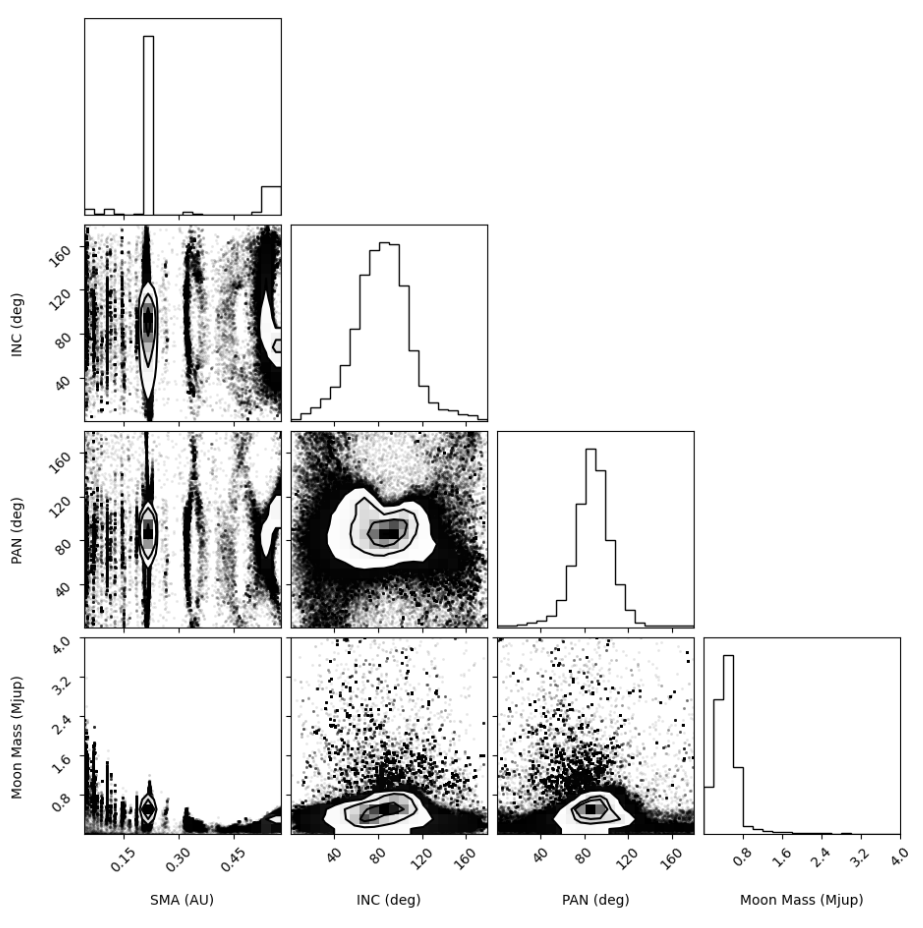}
   \caption{\label{figmcmcmoon} Fit of the astrometric data of HD 206893 B looking for an exomoon. The MCMC analysis finds a 95\% upper limit of 0.8 M$_{\rm Jup}$ for the potential exomoon mass and a peak in semi-major axis at 0.22 au, while the inclination and position angle of ascending node are around 90 deg. Another MCMC plot assuming a fixed semi-major axis around its peak is shown in Fig.~\ref{figmcmcmoonfixed}.}
\end{figure}

The posterior for the potential exomoon's semimajor axis (SMA), inclination (INC), position angle of ascending nodes (PAN), and mass are shown in Fig. \ref{figmcmcmoon}. Assuming no moon detection, we have a 95\% upper limit on the mass of the moon at 0.8 M$_{\rm Jup}$. However, we do find a peak in the moon SMA 1-D marginalized posteriors at about 0.22 au (period of 0.76 yr). The inclination (INC) and position angle of the ascending node (PAN) are moderately constrained, with best fitted with values around 90 deg, which would be $\sim$60 degrees misaligned from the orbit of B.

\begin{figure}[ht]
   \centering
   \includegraphics[width=9.cm]{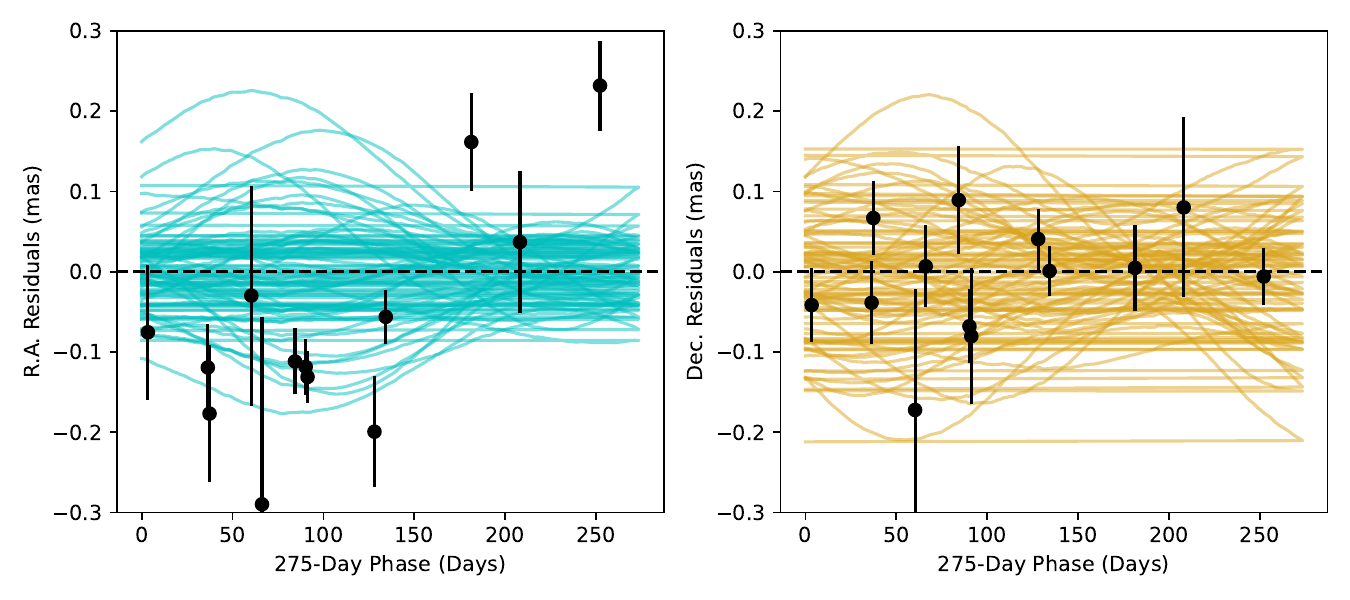}
   \includegraphics[width=9.cm]{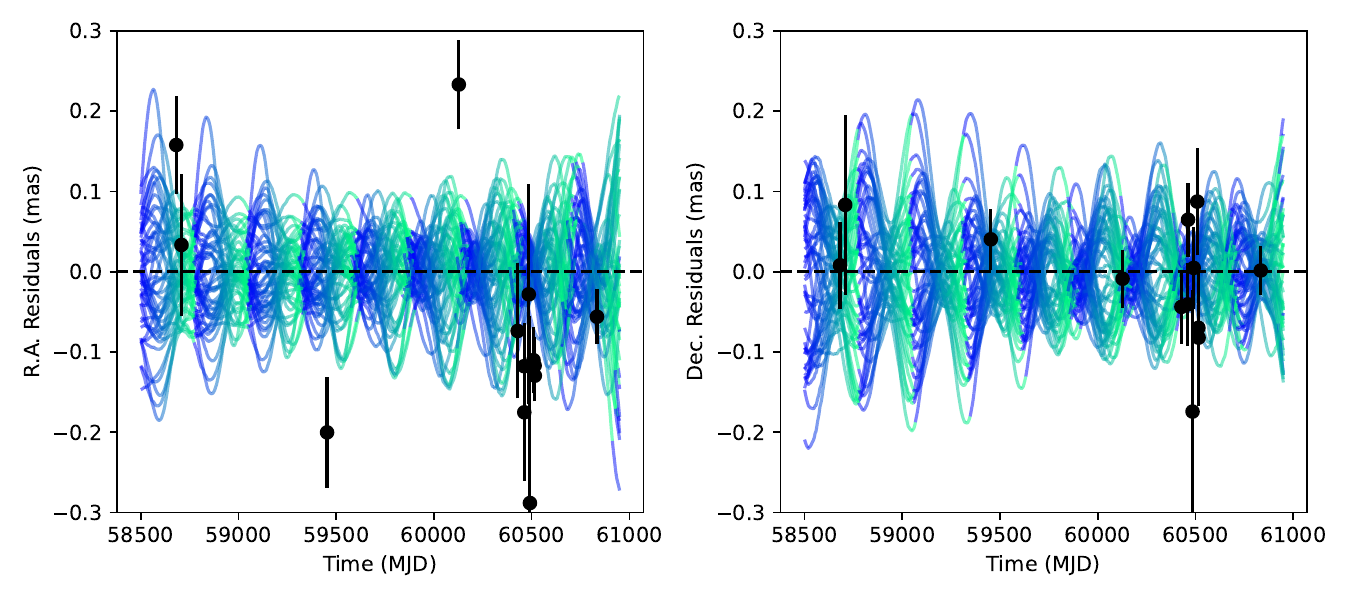}
   \caption{\label{figresidualsmoon} Residuals of HD 206893 B orbit after removing the average motion due to HD 206893 B orbiting the star and accounting for the motion induced by the planet c. In the top two plots, the data are phase-folded by a period of 275 days, using MJD 58500 as the starting point. 50 random draws with SMA between 0.21 and 0.26 AU are plotted in cyan for the RA and orange for the Dec Vs time plots, respectively. These model draws have the average motion of HD 206893 B removed, so they show both perturbation from the exomoon model and the residuals in the orbit of B from the average orbit of B.  In the bottom plots, the data are plotted as a function of MJD and not phase-folded. The orbits are colored by orbital phase, where blue is an orbital phase of 0 and green is an orbital phase of 1. Note that the orbital periods vary between each model drawn and are slightly different from an exact 275-day orbital period.}
\end{figure}

\begin{figure}[ht]
   \centering
   \includegraphics[width=9.cm]{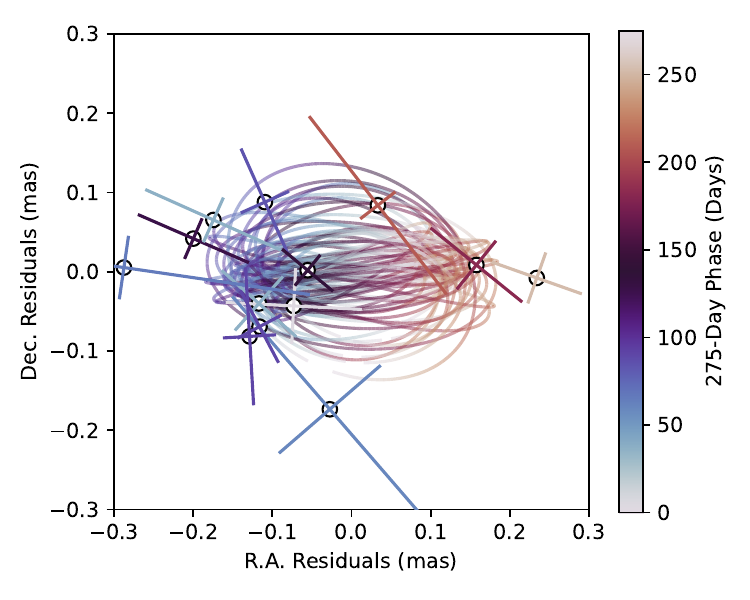}
   \caption{\label{figresidualsmoonb}  Similar residuals as Figure \ref{figresidualsmoon}, but projected onto the 2-D sky plane and the 50 random draws from the posterior only consider the motion of a potential exomoon (and not residual uncertainties in the orbit of B, which are shown in Figure \ref{figresidualsmoon}). The colors of both the data and the models correspond to the 275-day phase, as specified by the colorbar and defined in Figure \ref{figresidualsmoon}. }
\end{figure}

To analyze the most promising peak in the posterior, we consider only the posterior samples between 0.21 and 0.26 au. The posterior is plotted in Fig.~\ref{figmcmcmoonfixed}. In this case, the moon mass seems to converge to a value around 0.5 M$_{\rm Jup}$. We phase-fold the data based on this orbital period (0.76 yr) and compare it to the orbital model. The phase-folded residual plot shows the potential RA and Dec offsets on HD 206893 B after accounting for the planet c influence and removing the HD 206893 B orbit motion (Fig.~\ref{figresidualsmoon}). The offsets from 0 in the residuals are due to uncertainty in the orbit of HD 206893 B itself, while the sinusoidal wiggles are mainly caused by the moon in our orbital model. 
We also visualize the residuals on a RA Vs Dec plot in Fig.~\ref{figresidualsmoonb} (we also removed the average motion due to HD 206893 B orbiting the star and accounting for the motion induced by the planet c), where we see that the data seems to form an ellipse around the origin. The data seem to be well traced by exomoon trajectories from our model that are $\sim$0.15 mas in radius. We note that for a moon of mass 0.5 M$_{\rm Jup}$ orbiting at 0.22 au from B of mass 19.5 M$_{\rm Jup}$, then it causes the B component to orbit around a center of gravity that is located at about $0.5 \div 19.5 \times 0.22=0.0056$ au or $\sim$0.14 mas, which is indeed similar to what we observe on the RA Vs Dec plot. However, we note that systematic errors in the astrometry can also cause the data to form an ellipse around the origin. 

Systematic errors can have multiple origins. Table~1 in \citet{2014A&A...567A..75L} provides a review of their possible sources, including improper baseline calibration, terrestrial reference frame uncertainties, atmospheric dispersion, optical aberrations, and others. All of these effects were carefully evaluated and constrained to remain below 10\,$\mu$as. Laser trackers were used for baseline calibration; updates of the EOP (Earth Orientation Parameters) and the difference between UT1 (astronomical time) and UTC (coordinated universal time) are regularly obtained from the International Earth Rotation and Reference Systems Service (IERS); and optical aberrations within our system were measured and their impact assessed \citep{2021A&A...647A..59G}. Overall, we are confident that these known systematics do not dominate the error budget.

To test the hypothesis that the tentative detection might be real, we ran a simulation with no moons and compared it with the results of our previous fit, including a moon, by using the Bayesian information criterion (BIC). The median BIC favours the exomoon fit over the no moon fit by 7.4, which could point to some real signal but is less than the threshold of 10 to decisively favor the exomoon model \citep{Liddle2007}. In particular, the exomoon model could be used to fit for unaccounted systematics in our data. Therefore, our data do not allow us to confirm the presence of an exomoon for now, and further observations are needed. Furthermore, the BIC test does not factor into systematic errors in the astrometry, which the exomoon model could be fitting for.

We conclude that we find an upper limit of a potential moon around HD 206893 B of 0.8 M$_{\rm Jup}$ using the 95$^{\rm th}$ percentile limit or 0.5 M$_{\rm Jup}$ (or 9.3 M$_{\rm Nep}$), if we assume a tentative detection. We also see that the SMA is peaking at a value around 0.22 au, but we note that there are other smaller peaks around, and we cannot be certain that the period around 0.76 yr is real. More data from GRAVITY on a monthly timescale would allow us to conclude.
 
\section{Spectral analysis}
Together with the astrometric observation of HD 206893 B, we obtained an $R=4,000$ spectrum of the object. A previous spectrum at $R=500$ has already been published in \citet{2021A&A...652A..57K}, and in \citet{2025AJ....169..175S} at higher resolution ($R=35,000$). Moreover, SPHERE observations led to the detection of water at 1.4 microns \citep{2017A&A...608A..79D}.

\subsection{Atmospheric forward modeling}{\label{forward}}

We performed forward model fitting of the GRAVITY spectrum using the Bayesian framework Species \citep{2023ascl.soft07057S}, and several pre-computed atmospheric grids designed for young giant planets at medium spectral resolution: \texttt{Exo-REM} \citep{2018ApJ...854..172C,2021A&A...646A..15B}, \texttt{ATMO} \citep[grid from][]{2025A&A...701A.208P} and, for comparison, the commonly used \texttt{BT-Settl} grid \citep{allard_atmospheres_2012}.
These models are designed for late-M to early-T type objects and differ mainly in their treatment of clouds and thermal structure.
\texttt{BT-Settl} includes microphysical processes and settling to account for clouds; while  \texttt{Exo-REM} adds simple cloud microphysics and disequilibrium chemistry relevant for late L and early T dwarfs; \texttt{ATMO} is cloud-free model but incorporates diabatic processes that modify the temperature gradient via an effective adiabatic index $\gamma$.
For computational efficiency, we limited the fit to an effective temperature range of 1000–1900 K.
Results are in Figs. \ref{figatmapp_exorem}, \ref{figatmapp_atmo}, \ref{figatmapp_bt} for each model. Fig. \ref{figatmappspec} displays the best fit spectra together with the GRAVITY data).
The \texttt{BT-Settl} models yield higher effective temperatures and smaller radii, with $T_{\rm eff} = 1582^{+16}_{-29}$~K and $R = 0.93\,R_{\rm Jup}$, likely due to cloud modeling prescription..
In contrast, \texttt{Exo-REM} and \texttt{ATMO} provide broadly consistent effective temperatures, $T_{\rm eff} = 1162^{+158}_{-67}$~K and $T_{\rm eff} = 1097^{+60}_{-50}$~K, respectively, and larger radii, $R_{\rm p} = 1.79^{+0.24}_{-0.35}\,R_{\rm Jup}$ for \texttt{Exo-REM} and $R = 1.98^{+0.28}_{-0.30}\,R_{\rm Jup}$ for \texttt{ATMO}.
Models yield to the following C/O values, $0.38^{+0.20}_{-0.20}$ for \texttt{Exo-REM}, and $0.37^{+0.09}_{-0.05}$ for \texttt{ATMO}, consistent with sub-solar C/O ratios up to solar abundances for \texttt{Exo-REM} given the uncertainties. 

\subsection{Cross-correlation}\label{cross}
\begin{figure}[ht]
   \centering
   \includegraphics[width=9.cm]{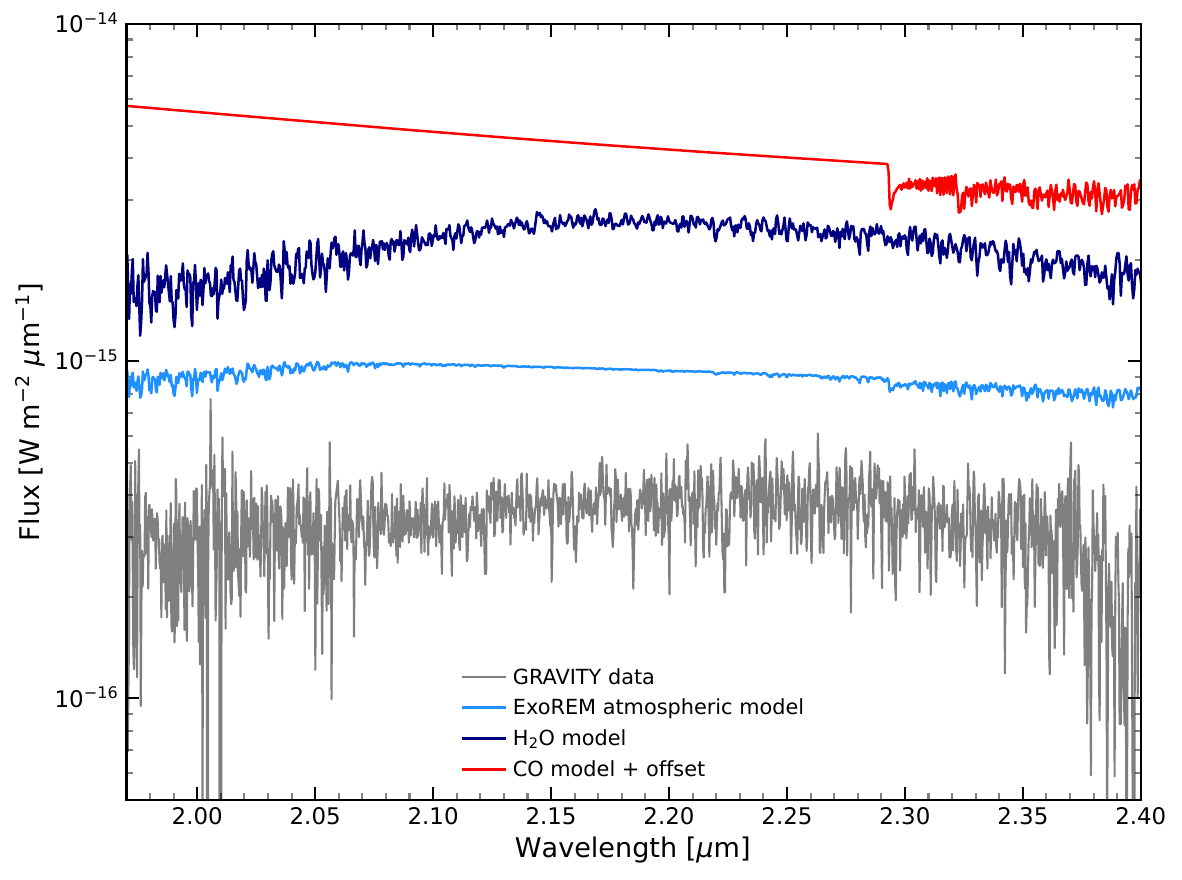}
   \caption{\label{figmodel} Comparison between the GRAVITY combined spectrum (gray) and the ExoREM synthetic spectrum for $T=1400$ K (light blue). Each species spectrum is shown with an offset with different colours – dark blue for water, pink for sodium, and red for CO.}
\end{figure}

Cross-correlation between model spectra and observed data is a commonly used technique to detect molecular species in the atmospheres of directly imaged exoplanets and brown dwarfs \citep[e.g.][]{2013Sci...339.1398K, 2024A&A...688A.116D}.
We cross-correlated the GRAVITY spectrum obtained in Fig.~\ref{figspectrum} with synthetic spectra generated using the ExoREM atmospheric model \citep{2018ApJ...854..172C,2021A&A...646A..15B}, assuming a range of effective temperature, $\log(g)=4.0$, solar metallicity, and a C/O ratio of 0.5. 
We found that an effective temperature of 1400 K provided the highest S/N, although the measured S/N remain similar across effective temperatures from 1100 to 1400 K. We focused on the main species expected to contribute in the K band spectral range, namely H$_2$O, CO, as predicted by the ExoREM model (see Figure \ref{figmodel}). While NH$_3$ and H$_2$S also exhibit spectral features, they are too faint to be detected in our data.

The observed spectrum was first smoothed using a Gaussian filter with $\sigma=10$ pixels. This smoothed spectrum is then subtracted to focus solely on the high-frequency components of the observed spectrum. 
The wavelength grid is sampled logarithmically, corresponding to a radial-velocity spacing of $\Delta v = 5~\mathrm{km~s^{-1}}$, and both the model and observed spectra are interpolated onto this new grid.
The CCF was then computed using the {\it scipy.signal.correlate} routine and normalized by the square root of the sum of the model and the data spectrum squared. We note that we haven’t included any broadening of the model given the low spectral resolution of the data.

\begin{figure}[ht]
   \centering
   \includegraphics[width=9.cm]{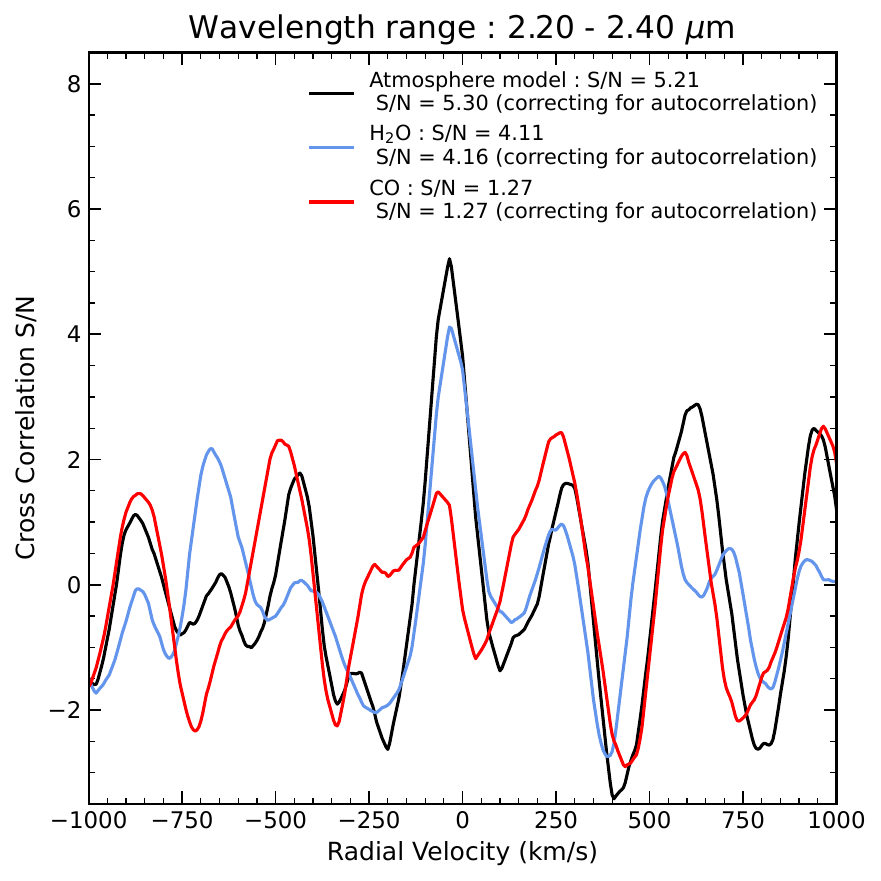}
   \caption{\label{figmodelccf} Cross-correlation S/N peaks between the data and the ExoREM synthetic spectrum for $T=1400$ K. The global cross-correlation S/N are computed as the peak divided by the standard deviation of the CCF wings (500 to 3000 km/s), and their final values are displayed in the legend for each species (in colour) and for the global atmospheric model (in black).}
\end{figure}

To account for the intrinsic autocorrelation of the model spectrum (which can be particularly strong for molecules with strong harmonic features such as CO), we also computed the autocorrelation function (ACF) of each model. This correction was performed by subtracting a rescaled ACF from the CCF wings ($\pm$ 200 km/s). Figure \ref{figmodelccf} displays each CCF along with the S/N of its peak.

The S/N is computed as the peak divided by the standard deviation of the CCF wings 
(between 500 km/s and 3000 km/s); 
we display the S/N with and without the autocorrelation corrected in the figure's legend. Restricting the analysis to the 2.2 – 2.4 ${\rm \mu}$m range, where water, CO features dominate, further improved the S/N of the detections. The full-model CCF yields a peak S/N of 5.3, with the H$_2$O-only template yielding an S/N of 4.16.
While no significant signal was detected for CO, the absorption band at 2.3 ${\rm \mu}$m appears to be visible in the spectrum (Fig.~\ref{figmodel}). The CCF peaks of H$_2$O and of the full model appeared around -35 km/s, though this offset remains below the instrumental resolution of GRAVITY of about 75 km/s. We note that the star radial velocity is -12.45 km/s \citep{2018yCat.1345....0G}, which is in the right range.

\section{Discussion}

\subsection{Target selection for GRAVITY+}\label{bestt}

Based on their respective distances, host star separations, and magnitudes, we filtered the Encyclopaedia of exoplanetary systems \citep[{\it exoplanet.eu},][]{2011A&A...532A..79S} for planets viable for a follow-up with GRAVITY+ aimed at detecting exomoons.
First, we selected confirmed planets and brown dwarfs exhibiting masses between 1 and 30 M$_{\rm Jup}$. Lower-mass objects would not be bright enough to be detectable with GRAVITY+. We also decided to remove free-floating planets because they are not yet accessible to GRAVITY+ without laser guide stars, as a bright reference source nearby is needed. Next, we retained only such planets with semi-major axes greater than 1 au and smaller than 100 au because they need to be well separated from the star to be detectable with GRAVITY+. Eccentricities were selected to be lower than 0.3 to avoid too complex dynamics in the potential moon systems. We also removed all planets detected around pulsars or other exotic objects. Finally, we selected only systems closer than 1000 pc from Earth, as astrometric signals from exomoons would become too small for too distant systems.

\begin{figure}
   \centering
   \includegraphics[width=9.cm]{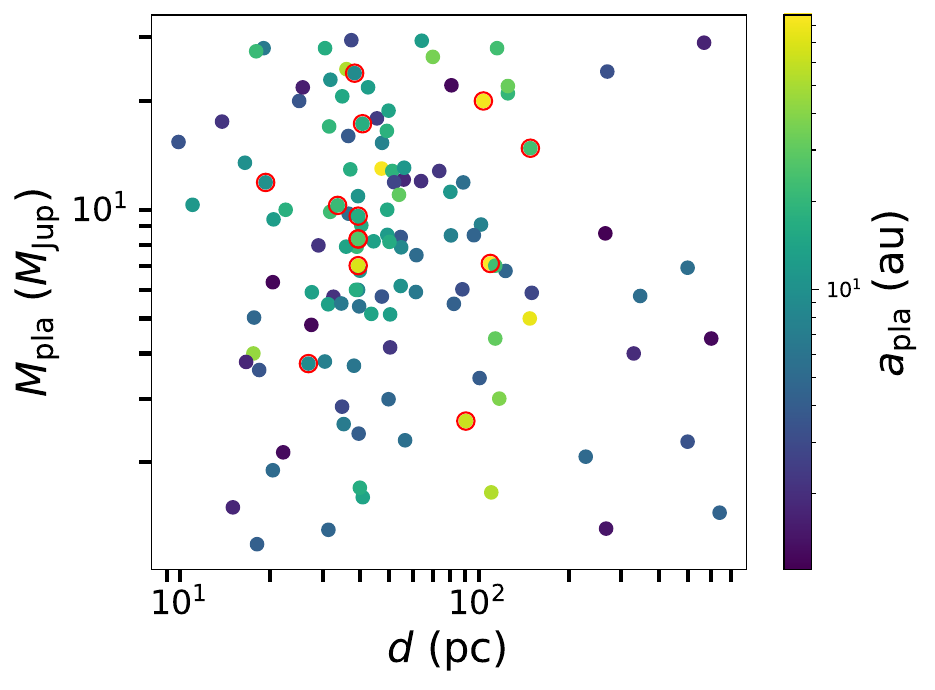}
   \caption{\label{figbesttargets} Selection of best targets filtered according to some specific criteria (see section \ref{bestt}) obtained from the exoplanet.eu database. The plot shows the planet mass $M_{\rm pla}$ Vs the distance $d$ for this selection of 124 objects, as well as the planet's semi-major axis $a_{\rm pla}$ in colour. The objects circled in red are those already observed with GRAVITY: AF Lep b, $\beta$ Pic b, c, HD 1160 b, HD 206893 B, HD 95086 b, HIP 65426 b, HIP 81208 Cb, HIP 99770 b, HR 2562 b, and HR 8799 b, c, d, e.}
\end{figure}

We show the resulting objects after this initial selection in Fig.~\ref{figbesttargets}, representing the planetary mass $M_{\rm pla}$ as a function of the target system distance $d$. The respective semi-major axis $a_{\rm pla}$ is shown in colour. At this stage, the sample consists of 124 planets. We circled in red the 14 objects that have already been observed with GRAVITY. Now, we selected the best 25 targets, i.e., with the smallest $d$ values, that have an angular distance greater than 0.08 arcsec (to be detectable with GRAVITY).

Using the \texttt{ATMO} \citep{2020A&A...637A..38P}, AMES-Dusty \citep{2000ApJ...542..464C, 2001ApJ...556..357A} and AMES-Cond \citep{2001ApJ...556..357A, 2003A&A...402..701B} evolutionary models, we computed the K-band magnitude expected for these most favourable targets. To this end, we used the available planet masses and system ages listed in {\it exoplanet.eu}, discarding targets for which these estimates were missing.
The resulting magnitudes are listed in Table~\ref{tab:model_magnitudes}.
Since the different evolutionary models are designed to handle a diverse set of objects, the resulting magnitude estimates are model-dependent. For instance, while the \texttt{ATMO} model lacks a cloud prescription, AMES-Dusty is capable of describing cloudy atmospheres.
No estimates were possible for objects with mass or age estimates outside the ranges in which the models are defined.
Some models could not be run when the evolutionary models are not defined in mass or age, leading to some empty values in Tab.~\ref{tab:model_magnitudes}.

In general, a GRAVITY+ follow-up requires a companion brighter than $K = 20$~mag. This narrows down the follow-up sample to 5 viable targets, namely, 2MASS J1315-2649 b \citep[full name is 2MASS J13153094-2649513 b,][]{2011ApJ...739...49B}, $\beta$ Pic b \citep{2010Sci...329...57L}, AF Lep b \citep{2023ApJ...950L..19F}, HD 155555 (AB) b \citep{2018A&A...619A..43A,2023NatCo..14.6232G}, and HD 60584 b, where the latter needs to be confirmed \citep{2022MNRAS.513.5588B}. Two of these, $\beta$ Pic b \citep{2020A&A...633A.110G} and AF Lep b \citep{2025AJ....169...30B}, were already observed with GRAVITY.
Finally, we retrieved the K-band fluxes of the individual hosts from the 2MASS All-Sky Catalogue \citep{2003yCat.2246....0C} and ensured that the host-to-companion contrast is within $10^{5}$ to be able to detect them with GRAVITY. All targets pass the contrast test, and this does not affect our previous list.

According to Eq.~\ref{Mmoon}, the moon mass detectable with GRAVITY scales as $T_{\rm moon}^{-2/3}$ (moon's period), $d$ (distance to Earth) and $M_{\rm pla}^{2/3}$ (planet's mass). Using that knowledge, we find that AF Lep b, HD 155555 (AB) b, and $\beta$ Pic b are the most favourable targets. The upper limits in mass for moons around HD 60584 b and 2MASS J1315-2649 b are higher by a factor of 3 to 4 since the companion hosting the moon is more massive. However, we note that other criteria may impact the presence of a massive moon. A recent study found that more massive moons are more likely to form around a super-massive Jupiter orbiting at 2 au around a solar-mass star \citep{2025A&A...699A.166D}, but more modelling is required to pinpoint the golden zone that would allow for the formation of the most massive moons around a variety of system architectures. 

For now, the five best targets – AF Lep b, HD 155555 (AB) b, $\beta$ Pic b, HD 60584 b (if confirmed), and 2MASS J1315-2649 b – should be considered as equally valid, as we cannot discard any of them as hosting a massive moon (i.e., a Neptune-mass moon or greater) that could be detectable with GRAVITY+. However, we note that for both HD 155555 (AB) b and the candidate HD 60584 b we would need more constraints from direct imaging to be able to track their orbits with GRAVITY. As for 2MASS J1315-2649 b, the host is a very low-mass star and it is too faint to use conventional adaptive optics. However, using the new laser guide stars may make it possible under the best atmospheric conditions, though it is at the limit \citep{2024arXiv240908438B}. For all those reasons, AF Lep b and $\beta$ Pic b appear to be the best short term exomoon targets. The main uncertainty is the moon's period, which could vary widely from system to system, thus only allowing for better upper limits for targets with moons at greater distances.
Indeed, if a system hosts a moon relatively far from its host planet, it will be easier to detect because it creates a stronger astrometric signal, and we will therefore be able to probe lower mass moons that may be more common.

\begin{table}[ht]
\centering
\caption{K-band magnitude estimates for some potential GRAVITY follow-up targets derived using different evolutionary models (\texttt{ATMO} \citep{2020A&A...637A..38P}, AMES-DUSTY \citep{2000ApJ...542..464C, 2001ApJ...556..357A} and AMES-COND \citep{2001ApJ...556..357A, 2003A&A...402..701B})}
\label{tab:model_magnitudes}
\begin{tabular}{lccc}
\toprule
Target name & \texttt{ATMO} & DUSTY & COND \\
\midrule
$\gamma$ Cephei A b     & 23.93 & {--}    & 26.80 \\
HD 128311 c         & 23.81 & 21.92 & 26.14 \\
GJ 504 b            & 20.35 & 22.11 & 21.16 \\
HD 89839 b          & 26.39 & 21.92 & 29.85 \\
HD 154345 b         & {--}    & {--}    & {--}    \\
2MASS J1315-2649 b  & 15.43 & 15.71 & 15.42 \\
$\beta$ Pic b          & 12.57 & 11.68 & 12.05 \\
HD 114783 c         & {--}    & {--}    & 53.62 \\
AF Lep b            & 15.87 & 16.82 & 15.51 \\
HD 155555 (AB) b     & 15.49 & 16.23 & 15.06 \\
HD 60584 b          & 15.63 & 16.04 & 15.66 \\
\bottomrule
\end{tabular}
\tablefoot{These filter magnitudes were calculated based on the \texttt{MKO\_K} (\texttt{ATMO}) and \texttt{Ks} (AMES-DUSTY and AMES-COND) filter transmission profiles.}
\end{table}

\subsection{The atmospheric makeup}

Using cross-correlation, we searched for the presence of different molecules and atoms expected to be present in the atmospheres of brown dwarfs and young giant exoplanets: H$_2$O, CO, CH$_4$, Na, K, HCN, CO$_2$, FeH, H$_2$S, HCN, NH$_3$, PH$_3$, TiO, and VO. For an effective temperature $\sim$1400K, atmospheric models predict spectra dominated by H$_2$O and CO in the K band. We obtain a relatively clear detection of H$_2$O by cross-correlation (SNR$\sim$4.1). Surprisingly, the signal is weak for CO (SNR$\sim$1.3). 
In addition to cross-correlation, we performed forward model fitting of the GRAVITY spectrum as described in section~\ref{forward}. 
The best fit (see Fig.~\ref{figatmapp_exorem}) corresponds to C/O=0.38(+0.2/-0.2), slightly sub-solar (but note that the posterior is bimodal). 
This is consistent with the non-detection of CO by cross-correlation. 

\subsection{Comparison with other methods to detect exomoons}

The search for exomoons presents significant challenges, and no confirmed detection exists to date. While the transit method \citep[e.g.,][]{2015ApJ...813...14K,2018AJ....155...36T} and radial velocity techniques \citep[e.g.,][]{2023AJ....165..113R} have driven most exomoon searches, they face inherent limitations. Transit techniques are sensitive primarily to moons on close-in orbits around transiting planets, require near-perfect orbital alignment, and disentangling moon signals from stellar activity or other planets can be complex. RV techniques require bright stars and struggle to target long-period moons \citep{2023AJ....165..113R}.

High-precision astrometry, as demonstrated in this work with VLTI/GRAVITY, offers a fundamentally complementary pathway. It directly measures the spatial wobble of the host planet or brown dwarf on the sky plane caused by an orbiting moon's gravitational pull. This technique possesses unique advantages as: 1) It can target planets with different inclinations, where the transit technique cannot, and the RV technique finds results that depend on the (most frequently) unknown $\sin(i)$ value.  2) It is ideally suited for searching for moons around directly imaged substellar companions, which typically reside on wide orbits that are more likely to host moons. Transiting planets have short orbital periods and thus small Hill spheres, which could be less favourable to moon retention \citep{2021PASP..133i4401D}, even if the planet formed further out and migrated inwards \citep{2016ApJ...817...18S}. 3) It is sensitive to wide-orbit moons with periods of months to years, which are inaccessible to transit techniques or current RV sensitivities \citep{2023AJ....165..113R}. Our tentative candidate around HD 206893 B exemplifies this, with a period of $\sim$ 0.76 years. The astrometry technique is even more powerful for moons at large distances from their planets, whereas the RV semi-amplitude decreases with larger separations, which could be complementary in some cases.

We note that the detection of circumplanetary discs could also help to better understand the formation sites of exomoons in the near future, which could prove very useful for testing models \citep[e.g., GQ Lup b,][]{2024ApJ...966L..21C}. If cavities are discovered (e.g., with the JWST) in these discs surrounding planets, this could indicate that moons are already forming at an early stage, and we could begin to study exomoon formation sites in detail, as well as the materials that accumulate as they form to forge these moons \citep[e.g.,][]{2024AJ....168..175H}.

\subsection[beta Pictoris b: Obliquity Constraints and Exomoon Predictions]{$\beta$ Pictoris b: Obliquity Constraints and Exomoon Predictions}

The young $\beta$ Pictoris system hosts two directly imaged super-Jupiters ($\beta$ Pic b and c), offering a prime laboratory for studying planetary dynamics. Upcoming JWST observations (programme GO 4758) will provide the first direct measurement of the spin-orbit obliquity of the outermost planet, $\beta$ Pic b — a milestone in exoplanet characterization.

Prior to JWST's definitive constraints, \citet{2024OJAp....7E.109P} combined dynamical and observational arguments to assess the likely obliquity of $\beta$ Pic b. Their analysis strongly favors a misaligned spin axis for the planet. Crucially, the authors propose that this misalignment could be maintained by a secular spin-orbit resonance driven by an unseen exomoon with a minimum mass exceeding Neptune's ($>1 M_{\rm Nep}$), and orbital period of 3–7 weeks.

Notably, such a companion may be accessible to astrometric detection: Depending on the moon mass, e.g., if it is rather massive and close to ten Neptune masses, the induced astrometric wobble of $\beta$ Pic b could fall within the capabilities of GRAVITY+. Given the system's brightness and proximity, we strongly advocate for rapid high-precision astrometric follow-up to test this prediction. A detection would not only validate the resonance mechanism but also provide the first direct evidence of an exomoon sculpting planetary obliquity.

\subsection{Misinterpreting a planet for a moon}

A fundamental challenge in astrometric exomoon detection is degeneracy: the reflex motion induced in a host planet (or brown dwarf) caused by an unseen planetary-mass companion orbiting the same central star can mimic the astrometric perturbation created by a moon. The astrometric displacement of a moon (see Eq.~\ref{eqast}) is roughly equal to $a_{\rm moon} M_{\rm moon}/M_{\rm pla}$. The signal from a planet would be roughly equal to $a_{\rm pla} M_{\rm pla}/M_{\star}$. 

Earlier, we found that the astrometric displacement is about 0.0056 au for our set of best-fit parameters for the moon. If it were created by a planet instead, it would mean that $a_{\rm pla} M_{\rm pla}/M_{\star} \lesssim 0.0056$ au. One can imagine a massive close-in planet would reproduce this signal, but then it may be detectable using RV. For instance, a planet with a period of 0.76 yr would correspond to a planet with a semi-major axis of about 0.9 au, which would imply a planet mass of $\sim$ 7 M$_{\rm Jup}$ to reproduce the astrometric displacement observed. Such planets can be detected in RV. A planet further out with a lower mass may also reproduce the signal in some systems with moons at different periods, but then its period will likely be too high compared to typical moons' periods, such as the one found here of 0.76 yr. These types of arguments should be checked on a case-by-case basis.

The fundamental difference between the effect of a moon and that of a planet is that the latter will cause a reflex motion on the star, while a moon would not. Hence, a follow-up of the star's RV would be important to confirm the detection if any doubts remain, e.g., the planet that may mimic the moon's signal is at the detection limit, or it is in a system that does not yet have RV limits or strong direct imaging constraints.

\section{Conclusions}
In this study, we presented the first dedicated high-precision astrometric search for an exomoon around a substellar companion, HD 206893 B, using VLTI/GRAVITY. Our analysis of the companion’s orbital motion revealed an upper limit of 0.8 M$_{\rm Jup}$ and a possible tentative astrometric residuals consistent with a perturbation potentially induced by an exomoon. If confirmed, this signal would correspond to a massive satellite ($\sim$0.5 M$_{\rm Jup}$) orbiting HD 206893 B with a period of approximately 0.76 years. We emphasize the tentative nature of this candidate, which needs further confirmation using GRAVITY data.

Beyond the exomoon signature, our work significantly refines the architecture of the HD 206893 system. Re-fitting the orbits of companions B and c suggests HD 206893 B resides on a wider orbit (10.75$\pm$0.08 au) than previously estimated, implying a lower mass for this object (19.5$^{+1.4}_{-1.3}$ M$_{\rm Jup}$). This revision may impact our understanding of the system’s formation and dynamical history. 

Complementing the astrometry, we analyzed the $R=4000$ spectrum of HD 206893 B obtained with GRAVITY, confirming a clear detection of water, and no CO detection. These spectral features provide crucial insights into the atmospheric composition and physical properties of this substellar companion.

Looking beyond HD 206893, we identify the most promising targets for future astrometric exomoon searches with GRAVITY. Based on proximity, companion brightness (K-band magnitude), orbital characteristics, and predicted astrometric signal strength, the five optimal candidates are: AF Lep b, HD 155555 (AB) b, $\beta$ Pic b, HD 60584 b, and 2MASS J1315-2649 b, with a priority to AF Lep b and $\beta$ Pic b that can be observed without further characterization of the orbits, which are sufficiently well known to follow them precisely with GRAVITY+.
These systems present the strongest prospects for detecting the tens of $\mu$as astrometric wobbles induced by exomoons within the current capabilities of VLTI/GRAVITY.

Our results demonstrate the potential of high-precision astrometry in the search for exomoons. While the detection around HD 206893 B requires further validation, this study establishes the methodology and proves the feasibility of the technique. GRAVITY, designed with the hope to reach micro-arcsecond precision \citep{2014A&A...567A..75L}, is currently the only instrument capable of pursuing this astrometric pathway to Neptune-like exomoons around directly imaged exoplanets and substellar companions. We therefore conclude that VLTI/GRAVITY has a pivotal role to play in the emerging field of exomoon and binary planet discovery. Future observations focusing on the HD 206893 system and the prioritized target list are essential to confirm the first exomoon and to usher in a new era of comparative exolunar science. New instruments at the VLTI with greater astrometric precision \citep[e.g.,][]{2025A&A...694A.277L} or a future kilometre baseline interferometric facility would allow reaching 1 $\mu$as \citep{2024arXiv241022063B}, which would push the detection limits more than an order of magnitude down.

\begin{acknowledgements}
This paper is dedicated to Rémi. We thank the referee for their thorough report that improved the paper substantially.
Based on observations collected at the European Southern Observatory under ESO programmes 105.20T0.001, 109.22ZA.002, 1103.B-0626, 1104.C-0651, 113.26D9.001 and 114.27UV.001.
This research has made use of the Jean-Marie Mariotti Center \texttt{Aspro}
service \footnote{Available at http://www.jmmc.fr/aspro}.
S.L.\ acknowledges the support of the French Agence Nationale de la Recherche (ANR-21-CE31-0017, ExoVLTI) and of the European Union (ERC Advanced Grant 101142746, PLANETES). P.K. acknowledges funding from the European Research Council (ERC) under the European Union's Horizon 2020 research and innovation program (project UniverScale, grant agreement 951549).
\end{acknowledgements}

\begin{appendix}

\begin{figure*}[ht]
\section{Spectral data at different epochs}
   \centering
   \includegraphics[width=18.cm]{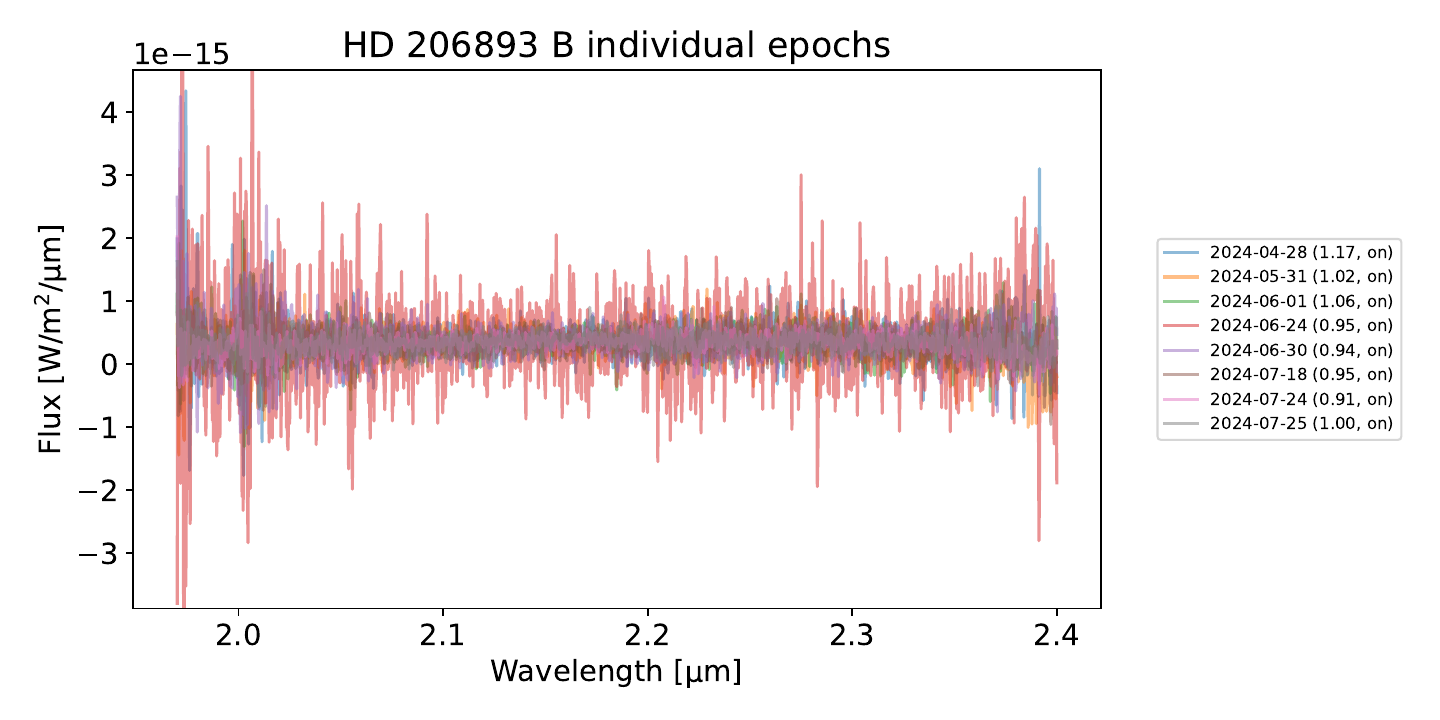}
   \caption{\label{figspectrumapp} Individual (color) spectra obtained with GRAVITY for HD 206893 B at the dates indicated in the figure's legend. The legend further reveals the best fitting scale factor for each epoch (see Section~\ref{sec:obs} for details) and that all epochs were observed in dual-field on-axis mode.}
\end{figure*}

\begin{figure}[ht]
\section{Orbital fit while fixing the semi-major axis of the moon}
   \centering
   \includegraphics[width=9.cm]{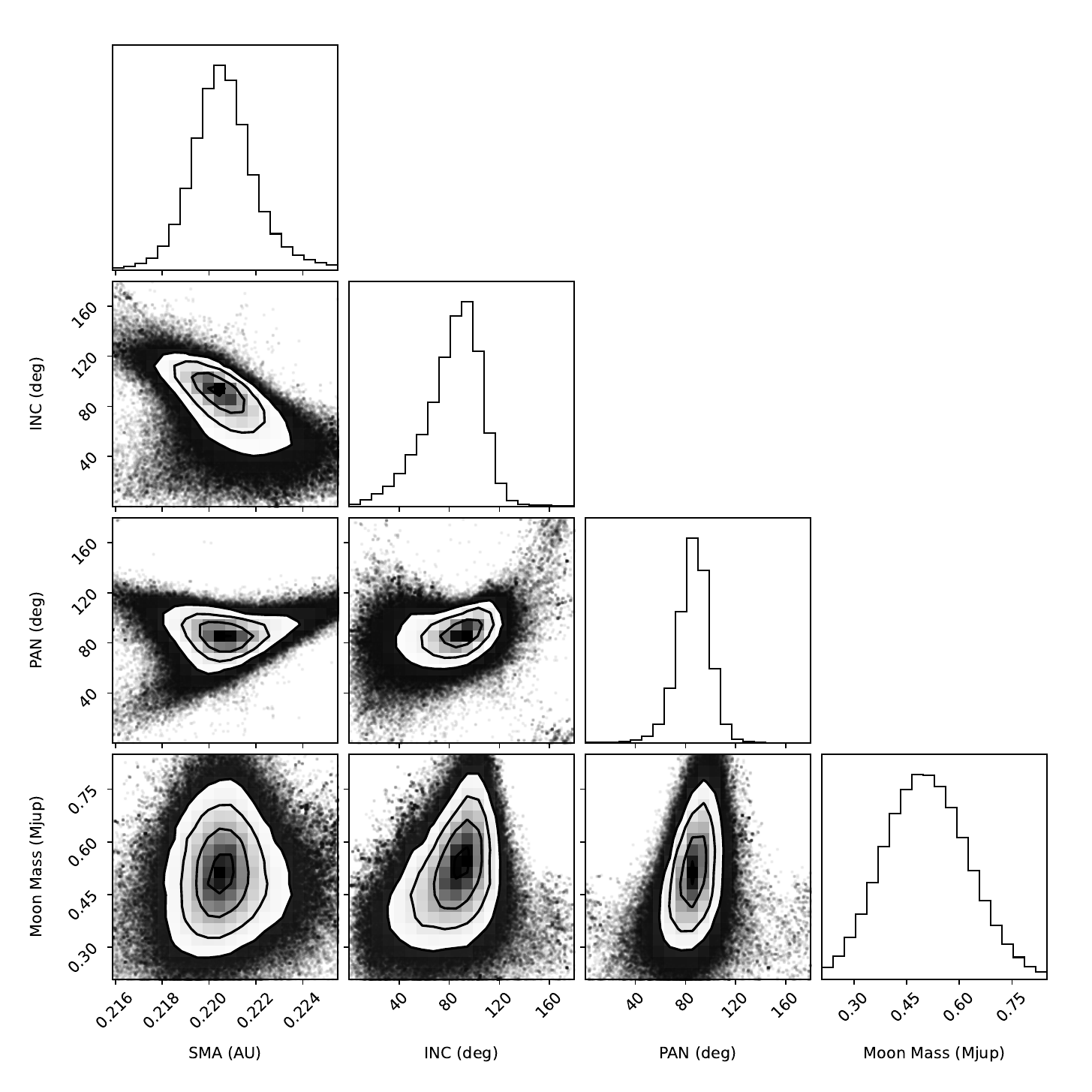}
   \caption{Fit of the astrometric data of HD 206893 B, looking for an exomoon when we limited the prior to be between 0.21 and 0.26 au. The MCMC analysis finds a tentative exomoon candidate of mass around 0.5 Jupiter masses.}
   \label{figmcmcmoonfixed}
\end{figure}

\FloatBarrier

\begin{figure}[ht]
\section{Bayesian analysis of the atmospheric spectrum of HD 206893 B}
   \centering
   \includegraphics[width=9.cm]{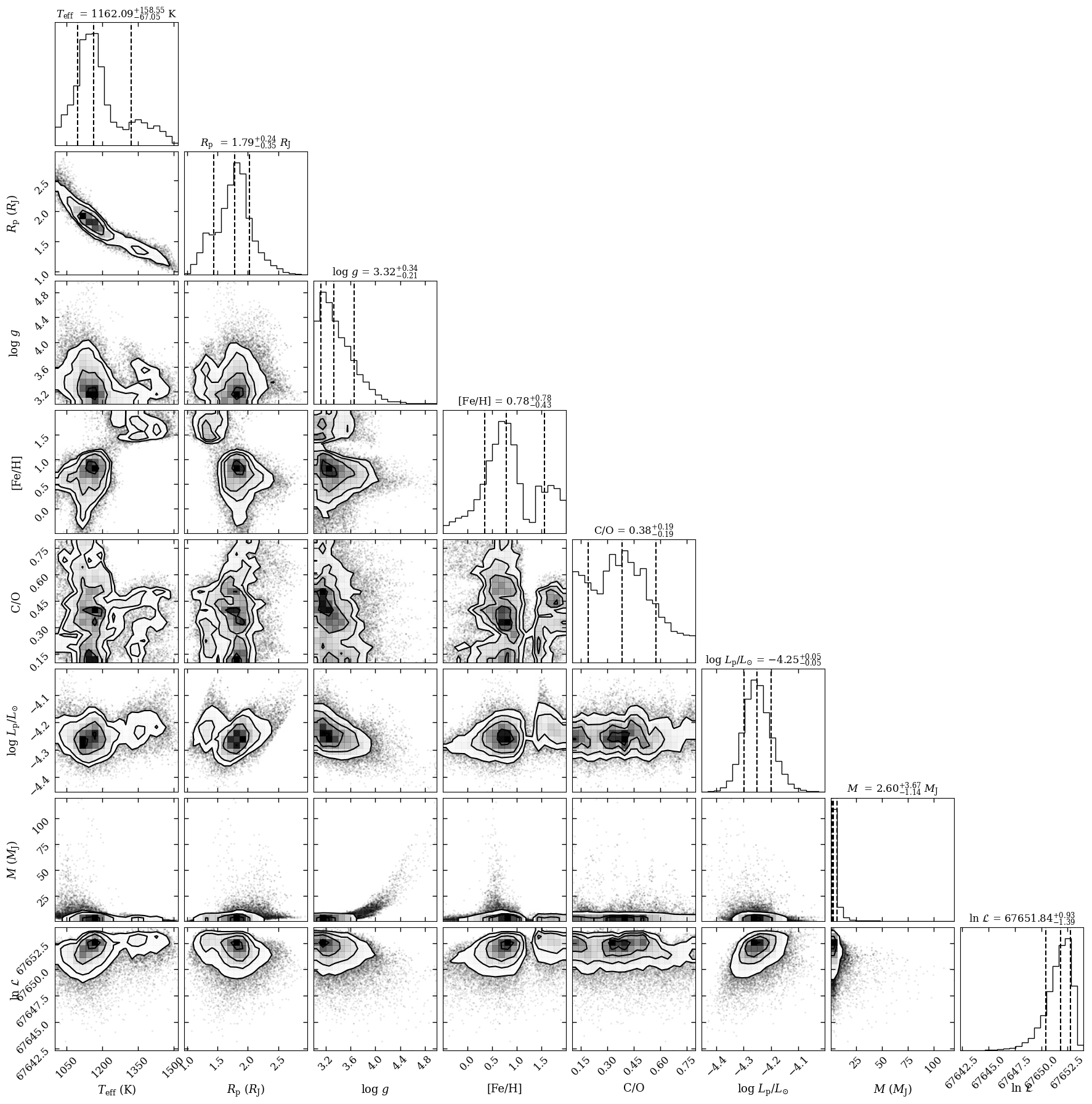}
   \caption{Posterior distributions obtained from Bayesian forward modeling of HD 206893 B’s atmospheric spectrum with ExoREM, displayed in a corner plot.}
   \label{figatmapp_exorem}
\end{figure}

\begin{figure}[ht]
   \centering
   \includegraphics[width=9.cm]{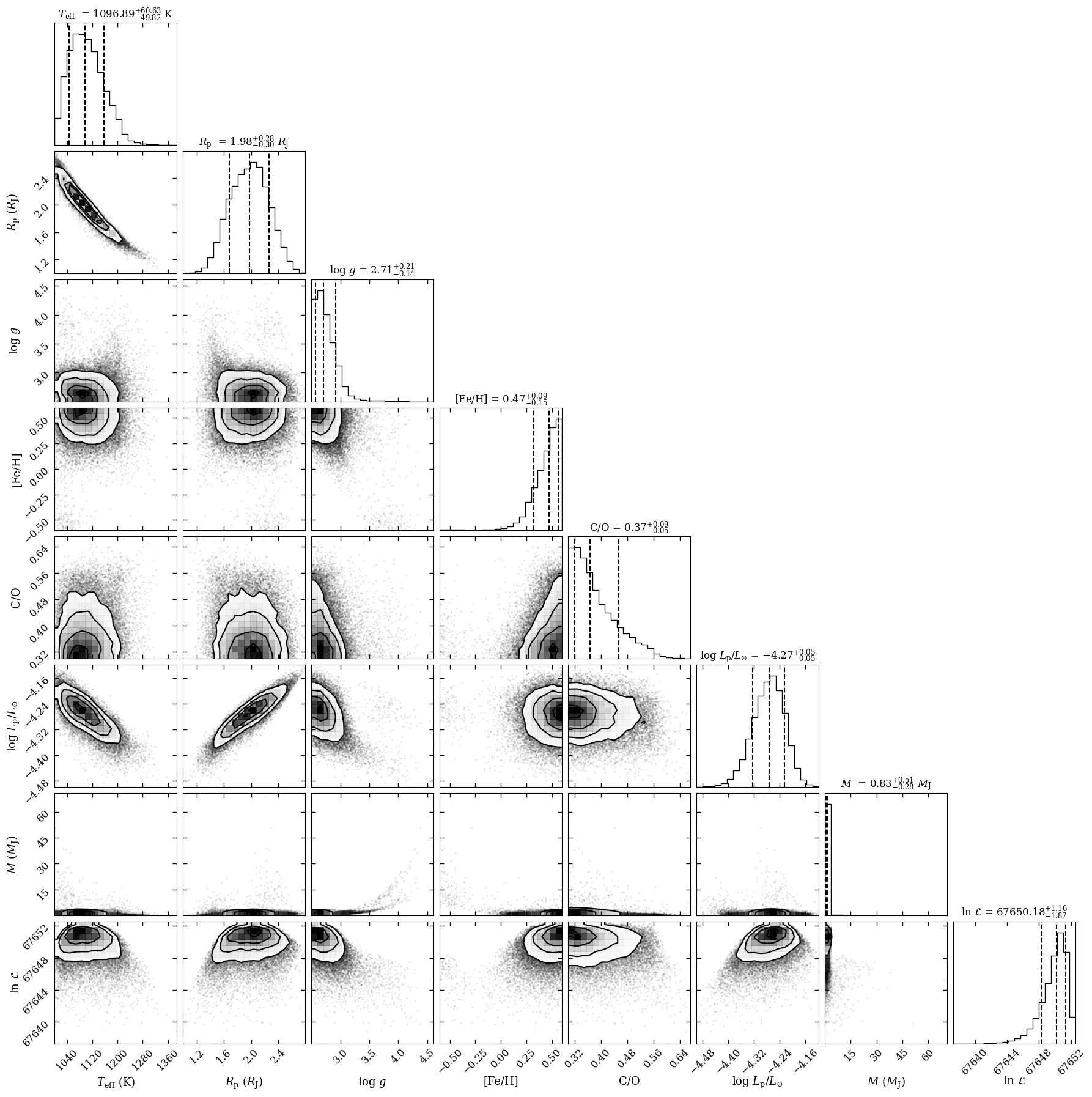}
   \caption{Posterior distributions obtained from Bayesian forward modeling of HD 206893 B’s atmospheric spectrum with ATMO, displayed in a corner plot.}
   \label{figatmapp_atmo}
\end{figure}

\begin{figure}[ht]
   \centering
   \includegraphics[width=9.cm]{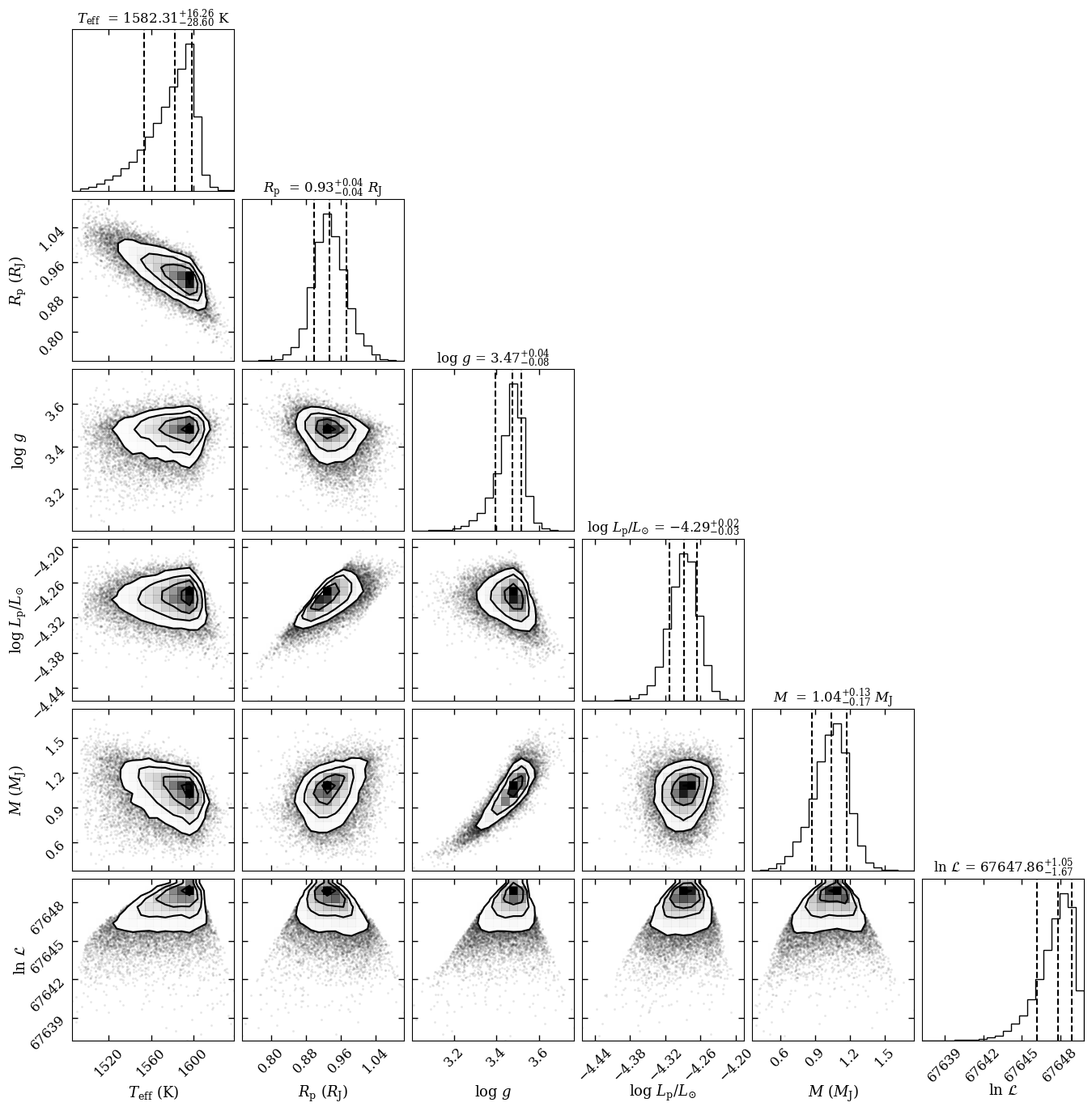}
   \caption{Posterior distributions obtained from Bayesian forward modeling of HD 206893 B’s atmospheric spectrum with BT-Settl, displayed in a corner plot.}
   \label{figatmapp_bt}
\end{figure}

\FloatBarrier

\begin{figure*}[ht]
   \centering
   \includegraphics[width=17.cm]{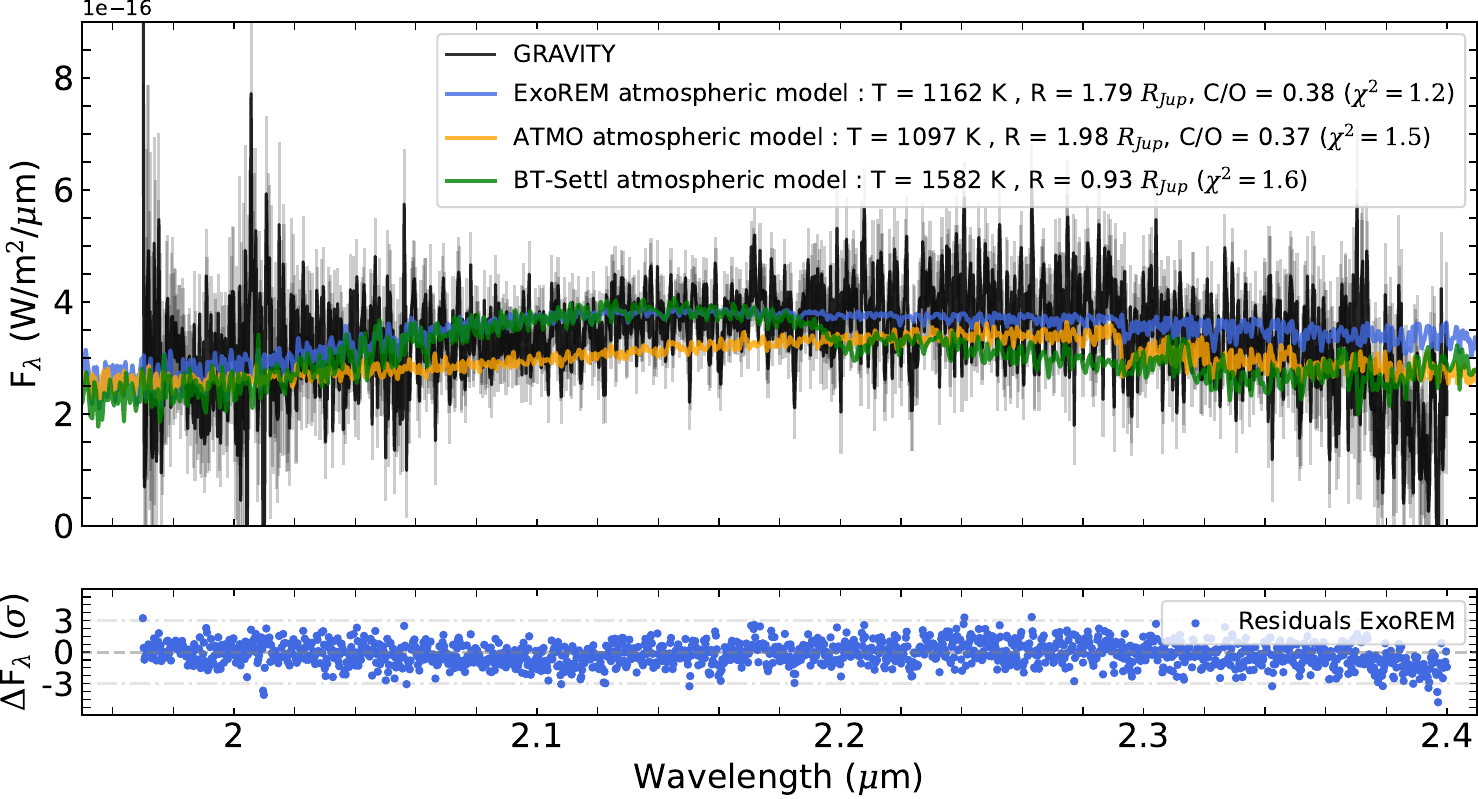}
   \caption{Best fit models overlaid on GRAVITY data for the forward modelling of the atmospheric spectrum of HD 206893 B using a Bayesian framework (upper panel). Residuals with ExoREM model are presented as an example (bottom panel).}
   \label{figatmappspec}
\end{figure*}

\FloatBarrier

\begin{sidewaystable*}
\section{Log of all GRAVITY observations}
\label{app:log}
\centering
\caption{Observation Log for HD206893}
\begin{tabular}{llccccccccc}
\hline
Object & Date & MJD & Start& End & DIT/NDIT/NEXP & airmass & $\tau_0$ & Seeing &  $\Delta_{\rm fiber}$RA & $\Delta_{\rm fiber}$DEC \\
 &  & (UT) & (UT) &  &  & (ms) & ('') & (mas) & (mas) \\
\hline
HD206893\,B & 2019-07-17 & 58681.392  & 08:52:56 & 09:56:06 & 60.0\,s / 12 / 5 & 1.2/1.5 &  1.5/2.2 & 1.11/1.84 &  132 & 199 \\
HD206893\,B & 2019-08-13 & 58708.160  & 03:21:16 & 04:21:09 & 60.0\,s / 12 / 5 & 1.0/1.1 &  3.1/5.3 & 0.73/1.18 &  132 & 199 \\
HD206893\,B & 2021-08-27 & 59453.093  & 01:18:42 & 03:08:23 & 30.0\,s / 16 / 4 & 1.0/1.3 &  3.6/6.3 & 0.49/0.83 &  17 & 201 \\
HD206893\,c & 2021-08-28 & 59454.125  & 02:12:21 & 03:49:18 & 10.0\,s / 32 / 5 & 1.0/1.1 &  1.7/7.0 & 0.55/0.90 &  -55 & -50 \\
HD206893\,c & 2021-09-28 & 59485.110  & 01:16:56 & 03:47:12 & 10.0\,s / 32 / 11 & 1.0/1.1 &  2.2/4.0 & 0.69/1.21 &  -76 & -83 \\
HD206893\,c & 2021-10-16 & 59504.061  & 23:45:31 & 03:07:57 & 10.0\,s / 32 / 27 & 1.0/1.3 &  1.6/3.7 & 0.45/0.65 &  -69 & -87 \\
HD206893\,c & 2022-05-22 & 59721.403  & 09:30:02 & 09:56:19 & 10.0\,s / 32 / 4 & 1.0/1.1 &  1.7/2.4 & 1.02/1.29 &  -32 & -95 \\
HD206893\,B & 2023-07-02 & 60127.218  & 05:04:49 & 05:24:04 & 30.0\,s / 8 / 4 & 1.2/1.3 &  4.1/6.4 & 0.35/0.46 &  -80 & 174 \\
HD206893\,B & 2024-04-28 & 60428.381  & 08:39:33 & 09:36:07 & 100.0\,s / 4 / 7 & 1.2/1.5 &  2.3/3.3 & 0.81/1.10 &  -120 & 148 \\
HD206893\,B & 2024-05-31 & 60461.388  & 09:07:57 & 09:32:33 & 100.0\,s / 4 / 3 & 1.0/1.0 &  2.6/6.0 & 0.52/0.71 &  -125 & 150 \\
HD206893\,B & 2024-06-01 & 60462.327  & 07:20:16 & 08:24:22 & 100.0\,s / 4 / 7 & 1.1/1.2 &  2.8/3.9 & 0.59/0.80 &  -125 & 150 \\
HD206893\,B & 2024-06-24 & 60485.334  & 07:30:10 & 08:33:51 & 100.0\,s / 4 / 7 & 1.0/1.0 &  2.3/2.3 & 1.37/1.93 &  -128 & 148 \\
HD206893\,B & 2024-06-30 & 60491.191  & 04:04:46 & 05:08:34 & 100.0\,s / 4 / 7 & 1.3/1.7 &  2.3/4.0 & 0.63/0.96 &  -128 & 148 \\
HD206893\,B & 2024-07-18 & 60509.364  & 08:07:14 & 09:23:36 & 100.0\,s / 4 / 8 & 1.1/1.3 &  2.8/5.9 & 0.70/1.44 &  -128 & 148 \\
HD206893\,B & 2024-07-24 & 60515.246  & 05:22:22 & 06:27:48 & 100.0\,s / 4 / 7 & 1.0/1.1 &  3.6/3.6 & 0.55/0.67 &  -128 & 148 \\
HD206893\,B & 2024-07-25 & 60516.263  & 05:47:28 & 06:51:15 & 100.0\,s / 4 / 7 & 1.0/1.0 &  2.8/4.1 & 0.55/0.72 &  -128 & 148 \\
HD206893\,B & 2025-06-08 & 60834.317  & 06:13:57 & 08:58:38 & 3.0\,s / 64 / 8 & 1.0/1.4 &  3.0/4.8 & 0.83/1.53 &  -170 & 112 \\
\hline
\end{tabular}
\label{tab:obslog}
\end{sidewaystable*}

\begin{table*}[ht]
\centering
\caption{Astrometry as a function of time and polynomial order}
\begin{tabular}{lccccc}
\hline
Date & MJD & N$_{\rm POLY}$ & RA & DEC & $\rho$\\
\hline
2019-07-17 & 58681.392 & 6 & $130.75 \pm 0.06$ & $198.13 \pm 0.05$ & -0.47 \\
  &   & 4 & $130.75 \pm 0.06$ & $198.11 \pm 0.05$ & -0.58 \\
  &   & 3 & $130.75 \pm 0.06$ & $198.11 \pm 0.05$ & -0.69 \\
2019-08-13 & 58708.160 & 6 & $127.06 \pm 0.07$ & $199.24 \pm 0.10$ & -0.90 \\
  &   & 4 & $127.12 \pm 0.08$ & $199.18 \pm 0.11$ & -0.92 \\
  &   & 3 & $127.12 \pm 0.09$ & $199.18 \pm 0.12$ & -0.95 \\
2021-08-27 & 59453.093 & 6 & $20.05 \pm 0.06$ & $205.83 \pm 0.03$ & -0.58 \\
  &   & 4 & $20.07 \pm 0.07$ & $205.83 \pm 0.04$ & -0.64 \\
  &   & 3 & $20.05 \pm 0.08$ & $205.83 \pm 0.04$ & -0.78 \\
2021-08-28\tablefootmark{a} & 59454.125 & 6 & $-76.52 \pm 0.08$ & $-82.71 \pm 0.12$ & -0.72 \\
  &   & 4 & $-76.56 \pm 0.14$ & $-82.61 \pm 0.10$ & -0.66 \\
  &   & 3 & $-5.21 \pm 0.36$ & $-42.76 \pm 0.26$ & -0.96 \\
2021-09-28 & 59485.110 & 6 & $-72.12 \pm 0.07$ & $-85.31 \pm 0.13$ & -0.85 \\
  &   & 4 & $-72.12 \pm 0.07$ & $-85.30 \pm 0.13$ & -0.86 \\
  &   & 3 & $-72.10 \pm 0.07$ & $-85.36 \pm 0.13$ & -0.82 \\
2021-10-16 & 59504.061 & 6 & $-69.32 \pm 0.04$ & $-86.73 \pm 0.07$ & -0.50 \\
  &   & 4 & $-69.32 \pm 0.06$ & $-86.73 \pm 0.07$ & -0.64 \\
  &   & 3 & $-69.28 \pm 0.05$ & $-86.73 \pm 0.07$ & -0.57 \\
2022-05-22 & 59721.403 & 6 & $-32.37 \pm 0.32$ & $-93.45 \pm 0.18$ & -0.97 \\
  &   & 4 & $-32.39 \pm 0.34$ & $-93.49 \pm 0.22$ & -0.91 \\
  &   & 3 & $-32.30 \pm 0.06$ & $-93.56 \pm 0.09$ & -0.06 \\
2023-07-02 & 60127.218 & 6 & $-79.30 \pm 0.05$ & $176.09 \pm 0.04$ & -0.34 \\
  &   & 4 & $-79.30 \pm 0.05$ & $176.06 \pm 0.04$ & -0.38 \\
  &   & 3 & $-79.28 \pm 0.06$ & $176.06 \pm 0.03$ & -0.37 \\
2024-04-28 & 60428.381 & 6 & $-120.95 \pm 0.08$ & $152.74 \pm 0.05$ & 0.03 \\
  &   & 4 & $-120.87 \pm 0.08$ & $152.72 \pm 0.04$ & 0.16 \\
  &   & 3 & $-120.92 \pm 0.06$ & $152.74 \pm 0.04$ & -0.09 \\
2024-05-31 & 60461.388 & 6 & $-125.28 \pm 0.05$ & $149.87 \pm 0.04$ & -0.18 \\
  &   & 4 & $-125.30 \pm 0.05$ & $149.82 \pm 0.04$ & -0.68 \\
  &   & 3 & $-125.30 \pm 0.06$ & $149.82 \pm 0.05$ & -0.73 \\
2024-06-01 & 60462.327 & 6 & $-125.45 \pm 0.08$ & $149.86 \pm 0.05$ & -0.70 \\
  &   & 4 & $-125.49 \pm 0.08$ & $149.86 \pm 0.04$ & -0.82 \\
  &   & 3 & $-125.47 \pm 0.08$ & $149.84 \pm 0.04$ & -0.82 \\
2024-06-24 & 60485.334 & 6 & $-128.30 \pm 0.18$ & $147.52 \pm 0.22$ & -0.73 \\
  &   & 4 & $-128.34 \pm 0.12$ & $147.56 \pm 0.11$ & -0.54 \\
  &   & 3 & $-128.30 \pm 0.09$ & $147.58 \pm 0.09$ & -0.71 \\
2024-06-30 & 60491.191 & 6 & $-129.38 \pm 0.21$ & $147.20 \pm 0.04$ & -0.50 \\
  &   & 4 & $-129.32 \pm 0.24$ & $147.20 \pm 0.05$ & -0.78 \\
  &   & 3 & $-129.29 \pm 0.23$ & $147.22 \pm 0.05$ & -0.81 \\
2024-07-18 & 60509.364 & 6 & $-131.50 \pm 0.04$ & $145.61 \pm 0.06$ & -0.64 \\
  &   & 4 & $-131.48 \pm 0.04$ & $145.61 \pm 0.06$ & -0.60 \\
  &   & 3 & $-131.50 \pm 0.04$ & $145.67 \pm 0.06$ & -0.58 \\
2024-07-24 & 60515.246 & 6 & $-132.27 \pm 0.03$ & $144.94 \pm 0.04$ & -0.07 \\
  &   & 4 & $-132.25 \pm 0.04$ & $144.92 \pm 0.04$ & -0.48 \\
  &   & 3 & $-132.27 \pm 0.03$ & $144.92 \pm 0.05$ & -0.63 \\
2024-07-25 & 60516.263 & 6 & $-132.40 \pm 0.03$ & $144.75 \pm 0.07$ & 0.16 \\
  &   & 4 & $-132.40 \pm 0.03$ & $144.86 \pm 0.04$ & -0.47 \\
  &   & 3 & $-132.40 \pm 0.03$ & $144.86 \pm 0.05$ & -0.57 \\
2025-06-08 & 60834.317 & 6 & $-170.50 \pm 0.04$ & $112.56 \pm 0.03$ & -0.53 \\
  &   & 4 & $-170.50 \pm 0.03$ & $112.56 \pm 0.03$ & -0.47 \\
  &   & 3 & $-170.50 \pm 0.03$ & $112.56 \pm 0.03$ & -0.45 \\
\hline
\end{tabular}
\tablefoot{
\tablefoottext{a}{First detection of HD206893\,c. The 3rd order polynomial fitting could not be used because of the faintness of the signal.}
}
\label{tab:pos}
\end{table*}

\end{appendix}

\label{lastpage}

\end{document}